\documentclass[11pt]{article}

\usepackage[margin=1in]{geometry}
\usepackage{newtxtext,newtxmath}
\usepackage{microtype}
\usepackage{setspace}
\usepackage{booktabs}
\usepackage{array}
\usepackage{graphicx}
\usepackage{caption}
\usepackage{float}
\usepackage{hyperref}
\usepackage{natbib}
\usepackage{fancyhdr}
\usepackage{tabularx}
\usepackage{makecell}
\usepackage{placeins}

\newif\ifdraft
\draftfalse   

\graphicspath{%
  {figures/}%
}

\newcommand{\CorpusOutcomeN}{175}
\newcommand{\CorpusStratumN}{8}
\newcommand{\EffectModeratePosN}{16}
\newcommand{\EffectModeratePosPct}{9.1}
\newcommand{\EffectStrongPosN}{5}
\newcommand{\EffectStrongPosPct}{2.9}
\newcommand{\EffectInconclusiveN}{109}
\newcommand{\EffectInconclusivePct}{62.3}
\newcommand{\EffectModerateNegN}{50}
\newcommand{\EffectModerateNegPct}{28.6}
\newcommand{\EffectStrongNegN}{16}
\newcommand{\EffectStrongNegPct}{9.1}
\newcommand{\BiasModeratePosN}{68}
\newcommand{\BiasModeratePosPct}{38.9}
\newcommand{\BiasStrongPosN}{33}
\newcommand{\BiasStrongPosPct}{18.9}
\newcommand{\HetModeratePosN}{86}
\newcommand{\HetModeratePosPct}{49.1}
\newcommand{\HetStrongPosN}{73}
\newcommand{\HetStrongPosPct}{41.7}
\newcommand{\JointBiasWeakEffectN}{45}
\newcommand{\JointBiasWeakEffectPct}{25.7}
\newcommand{\ClaimedEffectN}{108}
\newcommand{\ClaimedEffectPct}{61.7}
\newcommand{\ClaimedShrinkFiftyN}{82}
\newcommand{\ClaimedShrinkFiftyPct}{75.9}
\newcommand{\ClaimedEffectInconclusiveN}{86}
\newcommand{\ClaimedEffectInconclusivePct}{79.6}

\newcommand{\SIMnCells}{36}
\newcommand{\SIMnreps}{500}                  
\newcommand{\SIMnrepsdev}{150}               
\newcommand{\SIMnSyntheticOutcomes}{18{,}000}

\newcommand{\SIMB}{15{,}000}                 
\newcommand{\SIMBcheck}{5{,}000}             
\newcommand{\SIMBdev}{500}                   

\newcommand{\SIMMCSEdev}{0.0224}             
\newcommand{\SIMMCSEcheck}{0.0071}           
\newcommand{\SIMMCSEfinal}{0.0041}           

\hypersetup{
	pdftitle={Quantifying Evidential Rigor in Meta-Analytic Corpora: A Simulation-Characterized, Bias-Robust Bayesian Workflow},
	pdfauthor={Matt Hester},
	hidelinks
}

\title{\vspace{-0.5cm}\bfseries Quantifying Evidential Rigor in Meta-Analytic Corpora:\\[2pt] A Simulation-Characterized, Bias-Robust Bayesian Workflow with a Nutrition Case Study}

\author{
	Matt Hester\\
	Department of Mathematics and Statistics, University of Arkansas at Little Rock\\
	\texttt{mhester@ualr.edu}
}

\date{} 

\begin{document}
	\maketitle
	
	\vspace{-1.0em}
	\begin{center}
		{\footnotesize
			Companion repository:
			\href{https://github.com/matthewahester/evidential-audit-workflow}
			{\texttt{matthewahester/evidential-audit-workflow}}\\
			Archived release DOI:
			\href{https://doi.org/10.5281/zenodo.20467258}
			{\texttt{10.5281/zenodo.20467258}}
		}
	\end{center}
	\vspace{0.5em}
	
	\begin{abstract}
	Conventional meta-analysis typically summarizes evidence through pooled estimates, intervals, and \(p\)-values. These outputs do not directly measure how much evidence the data contain for an effect or for no effect, and evidence of publication selection or small-study effects is often treated as a diagnostic or sensitivity concern rather than integrated into the primary inference. Bayesian model averaging provides a natural structure for representing evidence as posterior weight over effect, heterogeneity, and bias-component model families.
	
	We introduce a corpus-scale, bias-robust Bayesian workflow for evidential auditing of meta-analytic corpora. The workflow reconstructs or accepts study-level effects and standard errors, harmonizes directions, fits a matched Bayesian random-effects baseline and a bias-aware model-averaged ensemble, and reports paired estimates with component and joint model-family evidence. Our implementation uses RoBMA-PSMA, but the framework applies more generally to ensembles exposing joint model-family weights.
	
	The central estimand is rigor: a joint Bayes-factor summary combining resolved effect/no-effect evidence with absence of an explicit bias component in the fitted ensemble. Rigor is not a positive-finding score; no-effect evidence can score highly, whereas inconclusive or bias-dependent evidence scores poorly.
	
	We characterize the workflow using an ADEMP-framed simulation/resampling design with known-cell synthetic simulation, empirical registry resampling, and empirical fitted-profile-weighted synthetic sampling. Nutrition intervention meta-analyses provide the worked case study, where bias-aware fitting often attenuates conventional estimates and many nominally meaningful effects lose clean evidential support. A public companion repository provides empirical inputs, generated empirical artifacts, simulation source/design grid, and documented workflow needed to inspect, reproduce, and adapt the audit.
\end{abstract}

\newpage
\section*{Highlights}

\subsection*{What is already known}

\begin{itemize}
	\item Conventional meta-analysis commonly reports pooled estimates, intervals, and \(p\)-values, but these summaries do not directly quantify evidence for an effect or evidence for no effect.
	\item Publication selection and small-study effects are often examined as diagnostics or sensitivity concerns, but are rarely integrated into the primary inferential summary.
	\item Bayesian model averaging provides a principled way to represent uncertainty across effect, heterogeneity, and bias-component model families.
\end{itemize}

\subsection*{What is new}

\begin{itemize}
	\item We introduce a corpus-scale evidential-audit workflow that fits a matched Bayesian random-effects baseline and a bias-aware model-averaged ensemble to harmonized meta-analytic outcomes.
	\item We define rigor as a single joint Bayes-factor estimand combining resolved effect/no-effect evidence with absence of an explicit bias component in the fitted ensemble.
	\item We characterize the workflow using an ADEMP-framed simulation/resampling design with known-cell synthetic simulation, empirical registry resampling, and empirical fitted-profile-weighted synthetic sampling.
	\item We provide a public companion repository with analysis-ready empirical inputs, generated empirical artifacts, simulation source and design grid, and documentation for adapting the workflow to new meta-analytic corpora.
\end{itemize}

\subsection*{Potential impact}

\begin{itemize}
	\item The workflow provides a portable way to compare evidential yield across strata, journals, intervention classes, research programs, or prospective evidence streams.
	\item Rigor does not reward positive findings alone: evidence for no effect can score highly, whereas inconclusive or bias-dependent evidence scores poorly.
	\item In a nutrition intervention case study, the workflow shows how bias-aware synthesis can attenuate conventional estimates and identify nominally meaningful effects that lack clean evidential support.
\end{itemize}

\noindent \textbf{Keywords:} Bayesian model averaging; evidential rigor; publication bias; meta-analysis; simulation studies

\section{Introduction}

Meta-analysis is central to cumulative science, but its standard summaries are not direct measurements of the evidence contained in the data. Pooled estimates, uncertainty intervals, and \(p\)-values describe estimation under a chosen model; they do not directly quantify how strongly the data support an effect, support no effect, or remain evidentially inconclusive. At the same time, threats such as publication selection, small-study effects, selective reporting, and study-level bias are widely recognized in evidence synthesis, but are often handled as diagnostics, sensitivity analyses, or qualitative judgments rather than incorporated into the primary inferential quantity. The result is a persistent separation between the reported meta-analytic conclusion and the evidential conditions that determine whether that conclusion should be considered reliable \citep{Ioannidis2005,Wasserstein2016,Wasserstein2019,McShane2019,Amrhein2017}.

This separation becomes more consequential at corpus scale. A single meta-analysis may leave unclear whether an apparent result reflects strong evidence, evidence for no effect, weak information, heterogeneity, or bias-structured reporting. A collection of meta-analyses raises a broader question: is effective evidence accumulation occurring? A journal, field, intervention class, funder portfolio, research program, or prospective evidence stream may produce many pooled estimates and many nominally meaningful results while generating little durable evidential resolution. Conversely, a research area can be evidentially productive when it efficiently rules out ineffective interventions. Corpus-level research synthesis therefore requires comparable measures of evidential yield, not only summaries of estimated effects.

Bayesian model averaging provides a natural structure for this problem. Rather than conditioning inference on one selected model, Bayesian model averaging assigns posterior probability to competing model families and expresses evidence as changes from prior to posterior model-family odds \citep{Hinne_2020,Gronau2021}. In meta-analysis, this structure allows effect presence, effect absence, heterogeneity, and bias-related components to be represented within a common inferential architecture. This is precisely the type of structure needed for an evidential audit: the evidence for a claim, the evidence for its absence, and the evidence for bias-relevant model components can be extracted from the same fitted ensemble.

Recent work in \emph{Research Synthesis Methods} has emphasized the importance of Bayesian workflow for bias adjustment in meta-analysis. Jung and Aloe demonstrate how prior predictive checks, posterior predictive checks, prior sensitivity analysis, model comparison, and targeted simulation can be used to evaluate a Bayesian study-level risk-of-bias adjustment model \citep{JungAloe2026}. Their contribution is important because it shows both the promise and difficulty of bias-adjusted Bayesian synthesis: bias adjustment can propagate uncertainty and yield more conservative estimates, but model behavior may be sensitive to prior assumptions, and predictive criteria may favor simpler random-effects models. Their framework is therefore best understood as a workflow for developing and validating a particular risk-of-bias-category adjustment model.

The present paper addresses a different but complementary problem. We do not propose a new study-level risk-of-bias model, nor do we attempt to construct bespoke priors for every individual meta-analysis. Instead, we take a documented bias-aware model-averaged ensemble as the audit engine and ask how its model-family evidence can be converted into a portable corpus-level measurement architecture. In the implementation used here, the audit engine is RoBMA-PSMA, which model-averages across effect, heterogeneity, publication-selection, and small-study-effect structures \citep{Maier2023,Bartos2023}. This choice is deliberate: a corpus-scale audit requires a common model space so that evidence measures are comparable across outcomes, strata, and corpora. A fully bespoke Bayesian workflow for each outcome would be difficult to scale and would undermine the comparability needed for evidential auditing.

The central estimand in this workflow is \emph{rigor}. Rigor is a single joint Bayes-factor summary that combines two requirements: resolved evidence for either an effect or no effect, and absence of an explicit bias component in the fitted ensemble. Rigor is not an adjusted-effect magnitude, a study-quality score, or a positive-finding score. An outcome can score highly because it provides evidence for an effect, or because it provides evidence for no effect. Outcomes with weak evidence, unstable direction, or conclusions that depend on publication-selection or small-study-effect components receive low rigor support. At corpus scale, distributions of rigor Bayes factors provide a way to compare whether evidence accumulation is effective: whether research activity is being converted into resolved evidence rather than merely into nominally significant pooled estimates.

The paper contributes three linked components. First, we define the rigor estimand and the associated component and joint model-family evidence summaries needed to compute it. Second, we specify a reproducible corpus-scale evidential-audit workflow: reconstructing or accepting study-level effect sizes and standard errors, harmonizing effect directions, fitting a matched Bayesian random-effects baseline and a bias-aware model-averaged ensemble, and reporting paired estimates together with evidence measures. Third, we characterize the workflow using an ADEMP-framed simulation and resampling design, following the simulation-study planning framework of \citet{Morris2019}. The simulation/resampling layer evaluates known-cell behavior, empirical registry stability, and empirical fitted-profile-weighted synthetic sampling. Its purpose is not to revalidate RoBMA-PSMA from first principles, but to evaluate whether the rigor estimand and audit workflow are stable and interpretable when the bias-aware ensemble is held fixed.

We demonstrate the workflow using a purposive corpus of nutrition intervention meta-analyses. Nutrition is a useful worked case study because the field is substantively important, intervention evidence is heterogeneous, and evidence bases often contain modest study counts, small-study patterns, and strong incentives for selective reporting \citep{Ioannidis2018,Brown_2023}. Modern nutrition evidence synthesis increasingly emphasizes structured risk-of-bias and certainty-of-evidence judgments, but those judgments are usually reported alongside rather than integrated into the primary quantitative synthesis \citep{Schwingshackl_2020,Stadelmaier_2024}. The nutrition corpus therefore provides a demanding demonstration of the audit workflow: it contains enough structure for stratum-level comparison, while also exhibiting the bias and heterogeneity concerns that motivate bias-aware evidence measurement. The case-study results are descriptive properties of the analyzed evidence objects, not population-level estimates for nutrition as a field.

The remainder of the paper proceeds as follows. Section~\ref{sec:methods} specifies the evidential-audit workflow as a methodological object, including corpus construction, reconstruction and standardization, matched baseline and bias-aware model fitting, diagnostic checks, computational implementation, and reproducible reporting. Section~\ref{sec:rigor-estimand} defines rigor as a joint model-family evidence estimand and describes how it can be summarized across outcomes, strata, and corpora. Section~\ref{sec:simulation} presents the ADEMP-framed simulation and resampling characterization. Section~\ref{sec:case-study} applies the workflow to the nutrition intervention case-study corpus. Section~\ref{sec:discussion} discusses portability, limitations, and future uses for corpus-level evidential auditing.

\section{Methods: a corpus-scale evidential-audit workflow}
\label{sec:methods}

\subsection{Workflow design and notation}
\label{sec:design}
\begin{figure}[H]
	\centering
	\includegraphics[width=\linewidth,height=0.82\textheight,keepaspectratio]{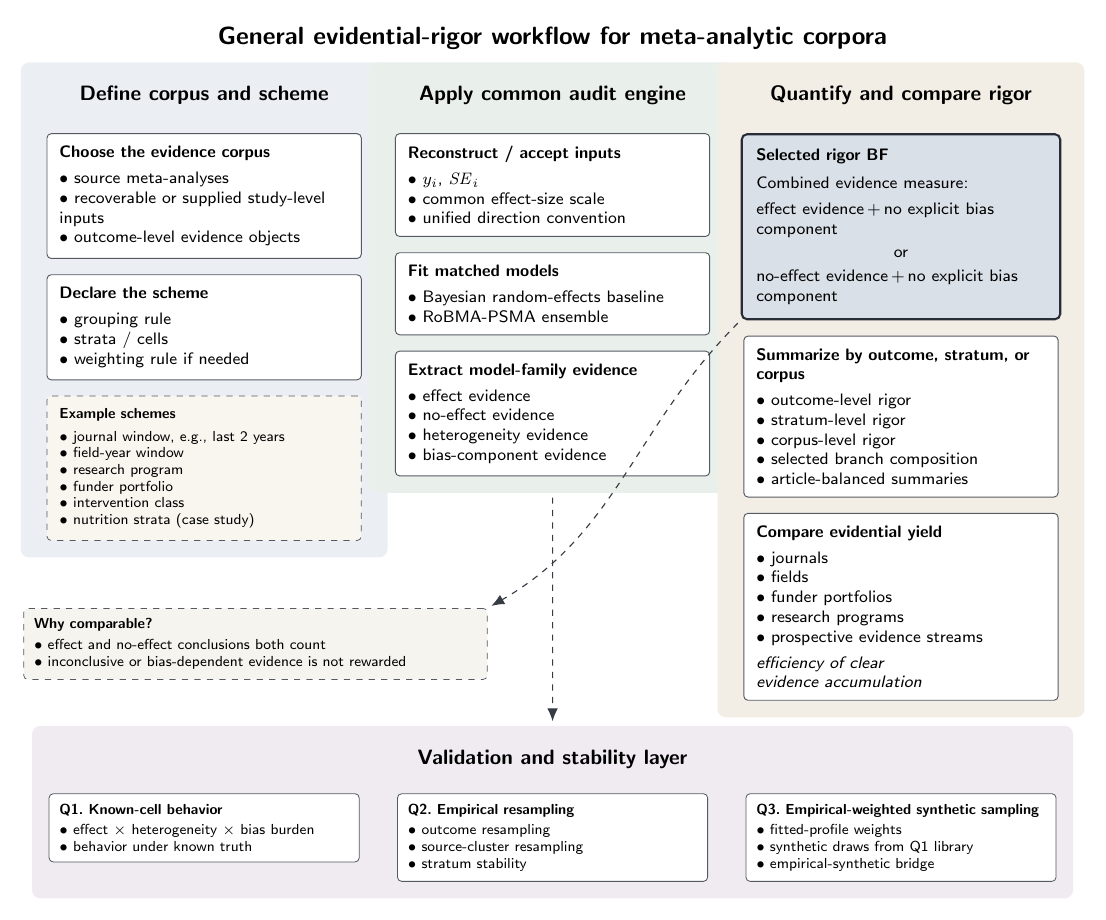}
	\caption{Corpus-scale evidential-rigor workflow.}
	\label{fig:rigor-workflow-architecture}
\end{figure}
We define the proposed method as a corpus-scale evidential-audit workflow. The workflow begins with a set of outcome-level meta-analytic evidence objects, applies a common Bayesian audit engine to each object, and then summarizes the resulting model-family evidence across analyst-declared groups. The goal is not to reproduce every source meta-analysis exactly, nor to construct a bespoke Bayesian model for every outcome. The goal is to place many reconstructed meta-analytic outcomes into a common evidential architecture so that effect evidence, no-effect evidence, heterogeneity evidence, and bias-component evidence can be compared across a corpus.

Figure~\ref{fig:rigor-workflow-architecture} gives the organizing map. Read left to right, the workflow first defines the evidence corpus and grouping scheme, then applies the same audit engine to every outcome-level evidence object, and finally summarizes rigor at outcome, stratum, and corpus scales. The lower layer of the figure anticipates the simulation/resampling characterization in Section~\ref{sec:simulation}; it is part of the method evaluation rather than a post hoc sensitivity appendix.

Let \(k=1,\ldots,n\) index reconstructed outcome-level meta-analyses in a corpus, and let \(i=1,\ldots,m_k\) index study-level estimates within outcome \(k\). The observed effect size for study \(i\) in outcome \(k\) is \(y_{ki}\), with standard error \(SE_{ki}\). The observed data object for outcome \(k\) is

\[
\mathcal{Y}_k
=
\{(y_{ki},SE_{ki}):i=1,\ldots,m_k\}.
\]

Here \(m_k\) denotes the number of study-level estimates inside one meta-analysis. This is distinct from the number of outcome-level meta-analyses summarized in a stratum or corpus-level draw, denoted later by \(n_{\mathrm{outcomes}}\) in the simulation/resampling section.

A \emph{scheme} \(S\) is an analyst-declared rule for grouping evidence objects. In the nutrition case study, the scheme is intervention domain, and its strata are caffeine, creatine, diet, fasting, fiber, omega-3, protein, and vitamin D. More generally, a scheme could group outcomes by journal window, field-year window, intervention class, funder portfolio, research program, source article, or prospective evidence stream. Let \(s_S(k)\) denote the stratum assigned to outcome \(k\) under scheme \(S\). The workflow is designed so that \(\mathcal{Y}_k\) is fit once and can then be summarized under any declared scheme without changing the underlying model fits.

At the outcome level, the audit engine returns a matched pair of analyses: a Bayesian random-effects baseline and a bias-aware Bayesian model-averaged ensemble fit to the same \(\mathcal{Y}_k\). At the reporting level, it returns paired posterior summaries, component model-family Bayes factors, joint model-family probabilities needed for rigor, and graphical diagnostics. At the corpus level, those outputs are reduced to stratum and corpus summaries, including outcome-weighted and article-balanced summaries when source articles contribute unequal numbers of reconstructed outcomes.

\subsection{Audit corpus, scheme, and scope of inference}
\label{sec:corpus}

An evidential audit begins by declaring a corpus of outcome-level meta-analytic evidence objects and a scheme for grouping those objects. The corpus determines which reconstructed or supplied meta-analytic datasets are analyzed; the scheme determines how those outcomes are summarized after fitting. Examples of possible schemes include intervention domains, journal-year windows, field-year windows, source articles, funder portfolios, research programs, or prospective evidence streams. The same fitted outcome registry can support multiple schemes when the relevant grouping variables are available.

The general workflow requires that each outcome-level evidence object can be represented by study-level effect sizes and standard errors on a common scale. It does not require that the source synthesis used the same model, software, estimator, or reporting convention. A source meta-analysis may enter the audit through an already-standardized dataset supplied by the analyst, through a published data file, or through reconstruction from reported study-level summaries. The audit begins only after the outcome has been reduced to an analysis-ready data object.

The nutrition corpus used in this paper is a worked empirical case study, not the definition of the method. It consists of a purposive set of published nutrition intervention meta-analyses selected because they were substantively visible, recoverable, and suitable for demonstrating the audit workflow across multiple strata. Corpus-level summaries should therefore be interpreted as descriptive properties of the analyzed audit corpus, not as population estimates for nutrition as a field.

The resulting demonstration corpus contains \CorpusOutcomeN{} reconstructed outcomes across eight intervention strata. The published source meta-analyses that define this corpus are included in the reference list for provenance; the machine-readable source registry and extraction records are provided in the companion repository. The same workflow could be applied to any declared corpus whose outcome-level meta-analytic datasets can be represented on the required effect-size and standard-error scale.

\subsection{Analysis-ready effect-size inputs and direction harmonization}
\label{sec:reconstruction}

For each outcome \(k\), the audit-ready input is a study-level dataset of effect-size estimates and standard errors on a common analysis scale. We write this generically as

\[
\mathcal{Y}_k
=
\{(y_{ki},SE_{ki}):i=1,\ldots,m_k\},
\]

where \(y_{ki}\) is the effect estimate for study \(i\) in outcome \(k\), \(SE_{ki}\) is its standard error, and \(m_k\) is the number of study-level estimates in that outcome. The matched Bayesian random-effects baseline, the RoBMA-PSMA ensemble, the rigor estimand, and the simulation/resampling layer are all defined from this common input structure. In the nutrition case study and simulation layer, the primary working scale is Hedges' \(g\), denoted \(g_{ki}\) with standard error \(SE_{g,ki}\) \citep{Hedges1981,Borenstein2009}. The more general workflow requires a common approximately normal effect-size scale with corresponding standard errors; it is not restricted to Hedges' \(g\).

The workflow can accept datasets that are already reported on the required scale, or it can use reconstructed inputs when source articles provide sufficient information to derive an effect estimate and standard error. Reconstruction is therefore an input-preparation step rather than a distinct inferential model. Its purpose is to create a common audit object for downstream fitting, not to reproduce every source paper's published pooled estimate. Source meta-analyses may differ in estimator choice, continuity corrections, repeated-measures conventions, software, effect-scale choices, and reporting granularity; the audit workflow standardizes the analysis-ready inputs and then applies the same baseline and bias-aware model ensemble to those inputs.

To make evidence comparable across heterogeneous outcomes, the workflow applies a unified direction convention before model fitting. Positive values are oriented toward the intervention-favored or substantively beneficial direction defined by the outcome context. When necessary, effects are multiplied by \(-1\) so that the favorable direction maps to the positive axis. This convention is especially important for bias-aware model averaging because one-sided selection components must be aligned with the declared effect direction. Direction harmonization does not force the posterior effect estimate to be positive and does not determine the selected rigor direction; it only defines the common coordinate system on which all outcomes are analyzed.

For the nutrition case study, source-specific reconstruction decisions, effect-scale conversions, repeated-measures assumptions, and exceptional sensitivity variants were documented in the extraction record. Those details are reported as case-study implementation details rather than as additional requirements of the general workflow.

\subsection{Matched Bayesian random-effects baseline}
\label{sec:baseline}

For each outcome \(k\), the unadjusted comparator was a Bayesian random-effects meta-analysis fit to the same audit-ready data object \(\mathcal{Y}_k=\{(y_{ki},SE_{ki}):i=1,\ldots,m_k\}\) used by the bias-aware ensemble. The baseline follows the standard normal--normal random-effects meta-analytic structure \citep{Borenstein2009,Gronau2021}. The baseline model is

\[
y_{ki}\mid \theta_{ki}
\sim
\mathcal{N}(\theta_{ki},SE_{ki}^2),
\qquad
\theta_{ki}\mid \mu_k,\tau_k
\sim
\mathcal{N}(\mu_k,\tau_k^2),
\]

so that marginally

\[
y_{ki}\mid \mu_k,\tau_k
\sim
\mathcal{N}(\mu_k,SE_{ki}^2+\tau_k^2).
\]

Here \(\mu_k\) is the mean effect for outcome \(k\) on the chosen analysis scale, and \(\tau_k\) is the between-study heterogeneity parameter on that same scale.

This baseline has a specific methodological role. It is not treated as the preferred final synthesis model and is not used to define rigor. Instead, it provides a standardized unadjusted reference fit against which the bias-aware model-averaged estimate can be compared. Directly comparing RoBMA-PSMA estimates to the source paper's reported pooled estimates would confound bias adjustment with differences in source-specific estimators, software, continuity corrections, effect-scale conventions, and random-effects implementations. By fitting the same Bayesian random-effects baseline to every reconstructed outcome, the workflow makes baseline-versus-bias-aware contrasts interpretable across the corpus.

The primary baseline-derived descriptive contrast is attenuation: the change in the estimated mean effect after moving from the matched random-effects baseline to the bias-aware RoBMA-PSMA ensemble. Attenuation is useful for diagnosing cases in which a conventional pooled effect is reduced by bias-aware modeling, but it is not the headline evidential-yield estimand. In particular, attenuation is not a natural success measure for corpora that primarily accumulate evidence for no effect. We therefore treat attenuation as a secondary audit metric that helps interpret the fitted evidence profile, while rigor provides the main evidence measure for comparing effect-supporting and no-effect-supporting outcomes on the same scale.
\subsection{Bias-aware Bayesian model averaging with RoBMA-PSMA}
\label{sec:robma}

The bias-aware component uses robust Bayesian meta-analysis with publication-selection model averaging (RoBMA-PSMA) \citep{Maier2023,Bartos2023}. The motivation is that the relevant meta-analytic structure is usually uncertain. An outcome may or may not contain a nonzero mean effect; between-study heterogeneity may be negligible or substantial; and the visible study record may or may not reflect publication-selection or small-study-effect structure. RoBMA-PSMA retains multiple candidate models for these possibilities and averages over them according to their posterior plausibility given the observed study-level effects and standard errors.

The base Bayesian model-averaged meta-analytic structure can be viewed as a crossed model space over effect and heterogeneity components \citep{Hinne_2020,Gronau2021},

\[
\mathcal{M}^{\mathrm{base}}_k
=
\mathcal{H}_{\mu,k}
\times
\mathcal{H}_{\tau,k},
\]

where

\[
\mathcal{H}_{\mu,k}
=
\{\mu_0,\mu_+\},
\qquad
\mathcal{H}_{\tau,k}
=
\{\tau_0,\tau_+\}.
\]

Here \(\mu_0\) denotes the effect-absent component, \(\mu_+\) the effect-present component, \(\tau_0\) the absence of between-study heterogeneity, and \(\tau_+\) the presence of heterogeneity. Crossing these components yields the familiar model-averaged meta-analytic structure that avoids a forced choice between null versus non-null and fixed-effect versus random-effects synthesis.

RoBMA-PSMA extends this structure by adding a bias-adjustment dimension,

\[
\mathcal{M}^{\mathrm{RoBMA}}_k
=
\mathcal{H}_{\mu,k}
\times
\mathcal{H}_{\tau,k}
\times
\mathcal{H}_{\omega,k}.
\]

The component \(\mathcal{H}_{\omega,k}\) contains models without an explicit bias-adjustment component, denoted \(\omega_0\), and models with explicit publication-selection or small-study-effect components, denoted \(\omega_+\). In the PSMA ensemble, \(\omega_+\) includes complementary bias-adjustment families, including selection models and PET-PEESE-type small-study-effect adjustments \citep{StanleyDoucouliagos2014}. Thus, the ensemble treats the form of bias adjustment itself as uncertain rather than requiring the analyst to select a single correction method in advance.

For outcome \(k\), let

\[
\mathcal{M}_k
=
\{M_{k1},\ldots,M_{kL}\}
\]

denote the resulting candidate model ensemble. Each model \(M_{k\ell}\) specifies a likelihood for \(\mathcal{Y}_k\), priors for the model-specific parameters, and one combination of the effect, heterogeneity, and bias-adjustment components above. Let \(\pi_{k\ell}=p(M_{k\ell})\) be the prior model probability and

\[
m_{k\ell}(\mathcal{Y}_k)
=
p(\mathcal{Y}_k\mid M_{k\ell})
\]

the marginal likelihood. The posterior model probability is

\[
p(M_{k\ell}\mid \mathcal{Y}_k)
=
\frac{
	\pi_{k\ell}m_{k\ell}(\mathcal{Y}_k)
}{
	\sum_{h=1}^{L}\pi_{kh}m_{kh}(\mathcal{Y}_k)
}.
\]

For any model-specific quantity \(\vartheta_k\), model-averaged inference is then obtained by weighting the model-specific posterior distributions by these posterior model probabilities:

\[
p(\vartheta_k\mid \mathcal{Y}_k)
=
\sum_{\ell=1}^{L}
p(\vartheta_k\mid \mathcal{Y}_k,M_{k\ell})
p(M_{k\ell}\mid \mathcal{Y}_k).
\]

This operation produces model-averaged posterior summaries for the mean effect, heterogeneity, and bias-aware effect estimates while propagating uncertainty over which model components are needed.

The same ensemble also yields model-family evidence, following the usual model-odds interpretation of Bayes factors and inclusion Bayes factors \citep{Hinne_2020,Gronau2021}. For any pre-specified model family \(\mathcal{A}_k\subset\mathcal{M}_k\), define the prior and posterior family probabilities

\[
p(\mathcal{A}_k)
=
\sum_{M_{k\ell}\in\mathcal{A}_k}p(M_{k\ell}),
\qquad
p(\mathcal{A}_k\mid\mathcal{Y}_k)
=
\sum_{M_{k\ell}\in\mathcal{A}_k}p(M_{k\ell}\mid\mathcal{Y}_k).
\]

The corresponding model-family Bayes factor is

\[
BF_{\mathcal{A}_k:\overline{\mathcal{A}}_k}
=
\frac{
	p(\mathcal{A}_k\mid \mathcal{Y}_k)/
	p(\overline{\mathcal{A}}_k\mid \mathcal{Y}_k)
}{
	p(\mathcal{A}_k)/
	p(\overline{\mathcal{A}}_k)
}.
\]

This Bayes factor measures how much the observed data change the odds for the model family \(\mathcal{A}_k\) against its complement. Component inclusion Bayes factors are obtained by choosing \(\mathcal{A}_k\) to be the family of models containing a given component, such as \(\mu_+\), \(\tau_+\), or \(\omega_+\). The corresponding component summaries quantify evidence for an effect, evidence for heterogeneity, and evidence for an explicit bias-adjustment component.

The notation \(\omega_0\) should be interpreted carefully. It denotes the subset of the fitted ensemble without an explicit publication-selection or small-study-effect component. It does not prove that the underlying literature is free of all bias. Similarly, evidence for \(\omega_+\) is evidence favoring the specified bias-adjustment components in the RoBMA-PSMA ensemble, not a complete diagnosis of every possible bias mechanism.

In this paper, RoBMA-PSMA is used as a fixed audit engine. This is a deliberate corpus-scale workflow choice. A common model space is needed so that model-family Bayes factors can be compared across outcomes, strata, and corpora. Constructing bespoke bias models or outcome-specific prior systems for every reconstructed meta-analysis would be difficult to scale and would weaken the comparability of the resulting evidence measures. The contribution of the present workflow is therefore not to redesign RoBMA-PSMA for each evidence object, but to apply a documented bias-aware ensemble consistently, extract component and joint model-family evidence, and use those quantities to define the rigor estimand in Section~\ref{sec:rigor-estimand}.

\subsection{Outcome registry and generated reporting contract}
\label{sec:reporting-layer}

After each matched baseline and RoBMA-PSMA fit, the workflow reduces the fitted objects to a standardized outcome-level sidecar. The sidecar is the fitted-model record for one outcome-level evidence object. It stores dataset identity, corpus and scheme metadata, posterior summaries from the matched baseline and bias-aware ensemble, component model-family Bayes factors, joint model-family probabilities needed for rigor, convergence and validation information, and references to diagnostic artifacts. The sidecar is not the manuscript reporting table; it is the auditable fitted-output record from which reporting is built.

This reduction is also a deliberate computational design. The full fitted model objects are large and expensive to regenerate, whereas the sidecars retain the outcome-level primitives needed for reporting, aggregation, and resampling. The workflow therefore separates expensive model fitting from downstream evidence summarization: each outcome is fit once under the accepted configuration, reduced to a versioned evidence record, and then reused for corpus summaries and simulation/resampling analyses. The resampling layer consequently evaluates variability in summaries of fitted evidence objects; it does not repeatedly refit RoBMA-PSMA inside each resampling draw.

After sidecar acceptance, the reporting layer rebuilds the accepted sidecars into a canonical outcome registry. This registry is the source for manuscript tables, corpus figures, simulation/resampling inputs, and aggregation summaries. Derived quantities such as attenuation, rigor direction, rigor category, article-balanced weights, and stratum-level summaries are regenerated from sidecar fields rather than entered by hand or recomputed independently by separate figure scripts. This keeps the estimand contract centralized while allowing the displayed tables and figures to update when extraction, fitting, or estimand definitions change.

Two levels of aggregation are reported by default. Outcome-weighted summaries treat each reconstructed intervention--outcome dataset as one evidence object. Article-balanced summaries downweight source articles that contribute many outcomes within the same stratum, preventing a prolific source synthesis from dominating a stratum summary solely through outcome multiplicity. These summaries answer different audit questions and are therefore reported together when outcome multiplicity is uneven.

The reporting layer writes both machine-readable CSV outputs and manuscript-ready \LaTeX{} fragments. This generated-artifact structure is part of the workflow: empirical results, simulation summaries, and manuscript displays are synchronized through the same registry and versioned estimand contract. The registry also preserves analysis-variant labels when a case-study-specific sensitivity analysis is required, allowing documented reconstruction or robustness variants to remain traceable without changing the general estimand definition, model-family evidence calculations, or aggregation rules.

\subsection{Registry-derived visual reporting layer}
\label{sec:diagnostic-artifacts}

The workflow includes a generated visual reporting layer downstream of the canonical outcome registry. This layer turns registry-derived quantities into standardized figures for inspecting outcome-, stratum-, and corpus-level evidence patterns. Because the figures are generated from accepted registry fields, they use the same definitions of rigor, attenuation, component evidence, direction labels, and article-balanced summaries as the manuscript tables. The visual layer therefore functions as a reproducible display system for the audit rather than as a separate source of estimand definitions.

The visual layer is organized at two scales. Stratum- and source-level figures support local inspection of a subset of the corpus, including paired baseline-versus-bias-aware effect displays, effect forests, component Bayes-factor distributions, and rigor rankings. Corpus-level figures support cross-stratum comparison, including component evidence distributions, attenuation summaries, rigor distributions, rigor direction and category composition, outcome-weighted versus article-balanced comparisons, and selected ranking displays. These figure families are generated from the same registry fields used for tables and summaries, so visual displays remain synchronized with the fitted evidence records and the versioned estimand contract.

Specific figures are interpreted in the sections where they answer substantive questions. Simulation and resampling figures are discussed in Section~\ref{sec:simulation}, where their mechanisms, sample-size axes, and performance measures are defined. Nutrition case-study figures are discussed in Section~\ref{sec:case-study}, where the corpus, strata, reconstruction issues, and descriptive findings are introduced. Per-outcome diagnostic artifacts, including \(z\)-plot diagnostics, are treated as case-study or supplementary inspection tools rather than as part of the general visual reporting contract.

\subsection{Computational reproducibility and companion repository}
\label{sec:computation}

The evidential-audit workflow is implemented in a public companion repository \citep{HesterEvidentialAuditWorkflow}:
\url{https://github.com/matthewahester/evidential-audit-workflow}. The repository is designed as a reusable implementation of the method, not merely as an archive of the nutrition case-study scripts. It contains the empirical nutrition audit, the simulation/resampling subproject, the shared estimand and schema layer, manuscript-facing generated outputs, and documentation for adapting the workflow to new meta-analytic corpora.

The implementation is organized around explicit contracts. The input contract specifies the analysis-ready study-level effect-size data required by the loader. The output contract specifies the fitted artifacts, sidecars, registry files, tables, figures, and audit products written by the active RoBMA 4.0 pipeline. A separate pipeline-layout document enforces the operating rule that each scheme/corpus uses its own output root, so empirical nutrition outputs and synthetic simulation outputs remain physically separate while sharing the same estimand and reporting machinery. This separation allows the same workflow to support the nutrition case study, the synthetic full36 simulation library, and future corpora without changing the core model-fitting or rigor-estimation code.

The script system is define-only. Sourcing a script installs functions, constants, and shared contracts, but does not fit models, read data, write files, or attach heavy runtime dependencies. Work is triggered only by explicit entry-point functions. The active empirical sequence is: discover and validate audit-ready datasets; fit the matched Bayesian random-effects baseline and RoBMA-PSMA ensemble; write outcome-level sidecars and audit artifacts; run sidecar acceptance checks; rebuild the canonical outcome registry; generate manuscript tables; and generate figures. The simulation/resampling layer uses the same main-pipeline registry and reducers after fitting synthetic audit-ready datasets under a separate output root.

This architecture separates expensive Bayesian computation from downstream evidence summarization. Full fitted model objects are large and may require substantial computation to regenerate. The sidecars and registry retain the primitive quantities needed for reporting---posterior summaries, component Bayes factors, joint model-family probabilities, rigor fields, diagnostic status, and identity metadata---so manuscript tables, corpus figures, article-balanced summaries, and resampling analyses can be rebuilt without rerunning MCMC. This also clarifies the target of the resampling layer: empirical and synthetic resampling operate on fitted outcome-level evidence records, not on repeated RoBMA-PSMA refits inside each resampling draw.

The sidecar acceptance gate is the reproducibility boundary between fitting and reporting. It verifies that sidecars use the current schema and estimand version, match the expected configuration, contain valid model-family evidence fields, and satisfy the rigor identities required by the reporting layer. A passing sidecar is structurally trustworthy for downstream reporting; scientific interpretation is based on the reported evidence quantities, diagnostics, and case-study context.

The repository is prepared for public release with a documented audit trail. Analysis-ready inputs and extraction records are retained as immutable inputs. Lean sidecars, audit CSVs, overview tables, generated \LaTeX{} fragments, and manuscript-facing figures are versioned as reproducibility primitives. Large fitted objects and other heavy binary artifacts may be omitted when they are regenerable from the committed inputs, scripts, sidecars, and recorded configuration. The working environment is documented separately, including verified versions of \textsf{R}, JAGS, \texttt{RoBMA}, \texttt{BayesTools}, and supporting packages. Additional documentation includes the data dictionary, output contract, script index, runbook, visual contract, diagnostic interpretation guide, release checklist, and artifact commit policy.

The main manuscript reports the inferential design, rigor estimand, simulation characterization, and nutrition case study. The companion repository provides the executable workflow needed to inspect, regenerate, and adapt the evidential audit.

\section{Rigor as a joint model-family evidential estimand}
\label{sec:rigor-estimand}

The previous section defines the audit workflow as a repeated application of a common bias-aware Bayesian model-averaged ensemble to outcome-level meta-analytic evidence objects. The remaining question is how the resulting model-family evidence should be reduced to a single outcome-level measure of evidential yield. Marginal component Bayes factors are useful, but none is sufficient by itself. A marginal effect Bayes factor can reward a positive finding even when the fitted ensemble also favors publication-selection or small-study-effect structure. A marginal no-bias Bayes factor can indicate support for models without an explicit bias component, but it does not say whether the evidence resolves toward an effect or toward no effect. Attenuation from the matched random-effects baseline to the bias-aware estimate is also informative, but it is an effect-size contrast rather than an evidence measure, and it is poorly suited to literatures where successful accumulation consists of ruling out ineffective interventions.

We therefore define \emph{rigor} as a joint model-family Bayes-factor estimand. Rigor measures whether the data update support toward one of two resolved, bias-component-absent branches of the fitted ensemble: an effect-supporting branch or a no-effect-supporting branch. The estimand is deliberately not a positive-finding score. Evidence for no effect can represent successful evidence accumulation, whereas weak evidence, unstable effect/no-effect resolution, or evidence that depends on explicit publication-selection or small-study-effect components receives little rigor support.

Throughout this section, absence of an explicit bias component has a model-space meaning. It refers to the subset of the fitted RoBMA-PSMA ensemble without the publication-selection or small-study-effect components included in that ensemble. It does not prove that the literature is free of all bias, does not certify internal validity, and does not encode study-level risk-of-bias domains unless those features are introduced through additional model components or metadata. Rigor is therefore evidence for a resolved effect/no-effect conclusion together with evidence against the need for the specified explicit bias-adjustment components.

\subsection{Why marginal component evidence is not enough}

RoBMA-PSMA reports marginal component evidence for the effect, heterogeneity, and bias-adjustment dimensions. These quantities remain important audit outputs, but each answers only a partial question. Let \(BF_k^\mu\) denote the marginal inclusion Bayes factor for the effect component and \(BF_k^\omega\) the marginal inclusion Bayes factor for explicit publication-selection or small-study-effect components. Then \(BF_k^\mu\) quantifies evidence for the effect-present component after averaging over heterogeneity and bias-adjustment structures, while \(BF_k^\omega\) quantifies evidence for explicit bias-adjustment components after averaging over effect and heterogeneity structures. The complementary no-bias Bayes factor is
\[
BF_k^{\bar{\omega}} = (BF_k^\omega)^{-1},
\qquad
\log_{10}BF_k^{\bar{\omega}}=-\log_{10}BF_k^\omega .
\]

These marginal summaries are not interchangeable with rigor. A large effect Bayes factor can occur in an outcome whose apparent effect is strongly entangled with explicit bias-adjustment structure. Conversely, a large no-bias Bayes factor can occur without resolving whether the accumulated evidence supports an effect or no effect. Component BFs therefore describe the fitted evidence profile, but they do not by themselves answer the corpus-audit question: how often does a research area produce resolved evidence that does not require an explicit publication-selection or small-study-effect component?

The matched baseline attenuation contrast answers a different question. It describes how the estimated mean effect changes when moving from the unadjusted Bayesian random-effects baseline to the bias-aware model-averaged ensemble. This is useful for identifying cases where conventional positive effects shrink under bias-aware fitting. It is not, however, a general evidential-yield metric. A mature literature that accumulates strong evidence for no effect may have little meaningful attenuation to report, yet still be evidentially successful. Rigor is introduced to measure this broader evidential success.

\subsection{Rigor branches in the RoBMA model space}

For outcome \(k\), let \(\mathcal{M}_k=\{M_{k1},\ldots,M_{kL}\}\) denote the RoBMA-PSMA model ensemble fit to the study-level evidence object
\[
\mathcal{Y}_k=\{(y_{ki},SE_{ki}):i=1,\ldots,m_k\}.
\]
As in Section 2.5, the top-level model dimensions are the effect component \(\mu\), the heterogeneity component \(\tau\), and the explicit bias-adjustment component \(\omega\). We write
\[
H_{\mu,k}=\{\mu_0,\mu_+\},\qquad
H_{\tau,k}=\{\tau_0,\tau_+\},\qquad
H_{\omega,k}=\{\omega_0,\omega_+\}.
\]
Here \(\mu_+\) denotes the effect-present component, \(\mu_0\) the no-effect component, \(\tau_+\) and \(\tau_0\) denote the presence or absence of between-study heterogeneity, and \(\omega_+\) denotes models with explicit publication-selection or small-study-effect components. The family \(\omega_0\) denotes models without those explicit bias-adjustment components in the fitted ensemble.

Rigor is defined from two pre-specified branches. The effect-supporting branch is the subset of models with the effect component present and no explicit bias component, marginalizing over heterogeneity:
\[
\mathcal{R}_k^{+}
=
\mathcal{M}_{k,\mu_+\cdot\omega_0}
=
\mathcal{M}_{k,\mu_+\tau_0\omega_0}
\cup
\mathcal{M}_{k,\mu_+\tau_+\omega_0}.
\]
The no-effect-supporting branch is the subset of models with the no-effect component and no explicit bias component, again marginalizing over heterogeneity:
\[
\mathcal{R}_k^{0}
=
\mathcal{M}_{k,\mu_0\cdot\omega_0}
=
\mathcal{M}_{k,\mu_0\tau_0\omega_0}
\cup
\mathcal{M}_{k,\mu_0\tau_+\omega_0}.
\]
The dot in \(\mathcal{M}_{k,\mu_\cdot\cdot\omega_0}\) indicates that the heterogeneity dimension is not fixed. Heterogeneity is treated as an important fitted component, but not as a disqualifying feature for rigor: both fixed-effect-like and heterogeneous evidence can be rigorous if the evidence resolves toward either effect or no effect without requiring an explicit bias-adjustment component.

\subsection{Branch-specific and headline rigor Bayes factors}

For any pre-specified model family \(\mathcal{A}_k\subset\mathcal{M}_k\), Section~\ref{sec:robma} defines the model-family Bayes factor
\[
BF_{\mathcal{A}_k:\overline{\mathcal{A}}_k}
=
\frac{
	p(\mathcal{A}_k\mid\mathcal{Y}_k)/p(\overline{\mathcal{A}}_k\mid\mathcal{Y}_k)
}{
	p(\mathcal{A}_k)/p(\overline{\mathcal{A}}_k)
}.
\]
Applying this definition to the two rigor branches gives
\[
BF_k^{R,+}
=
BF_{\mathcal{R}_k^{+}:\overline{\mathcal{R}_k^{+}}},
\qquad
BF_k^{R,0}
=
BF_{\mathcal{R}_k^{0}:\overline{\mathcal{R}_k^{0}}}.
\]
The first quantity is evidence for the effect-supporting no-bias branch; the second is evidence for the no-effect-supporting no-bias branch.

The headline rigor Bayes factor is the stronger of these two branch-specific Bayes factors on the log scale:
\[
\log_{10}BF_k^R
=
\max\{\log_{10}BF_k^{R,+},\log_{10}BF_k^{R,0}\}.
\]
The selected branch is
\[
b_k^R
=
\arg\max_{b\in\{+,0\}}
\log_{10}BF_k^{R,b}.
\]
Here \(b_k^R=+\) denotes the effect branch and \(b_k^R=0\) denotes the no-effect branch. In the machine-readable output this branch label is stored as \texttt{rigor\_direction}, with values \texttt{effect} and \texttt{no\_effect}.

The selected branch label should not be interpreted without the magnitude and sign of \(\log_{10}BF_k^R\). When \(\log_{10}BF_k^R\) is positive and substantively large, the selected branch identifies whether the outcome provides clean effect-supporting evidence or clean no-effect-supporting evidence. When \(\log_{10}BF_k^R\) is near zero, the outcome provides little evidence for either clean resolved branch. When \(\log_{10}BF_k^R\) is negative, even the better-supported clean branch has lost support relative to its complement. Such cases should be interpreted as failures of clean evidential resolution, not as evidence for the selected branch.

\subsection{Relationship to marginal component Bayes factors}

Rigor is a joint model-family estimand. It is not generally equal to a product of marginal component Bayes factors. In particular,
\[
BF_k^{R,+}
\neq
BF_k^{\mu_+}\,BF_k^{\bar{\omega}},
\qquad
BF_k^{R,0}
\neq
BF_k^{\mu_0}\,BF_k^{\bar{\omega}}
\]
in general, because the marginal effect/no-effect and no-bias families average over different partitions of the model space. Exact rigor extraction requires the prior and posterior probability mass of the joint branches \(\mathcal{R}_k^+\) and \(\mathcal{R}_k^0\), obtained by summing over the individual model combinations in the fitted ensemble.

The headline rigor Bayes factor is also not the Bayes factor for the union
\[
\mathcal{R}_k^+\cup\mathcal{R}_k^0.
\]
That union collapses to the full \(\omega_0\) family,
\[
\mathcal{R}_k^+\cup\mathcal{R}_k^0
=
\mathcal{M}_{k,\cdot\cdot\omega_0},
\]
because it includes both the effect-present and no-effect branches while marginalizing over heterogeneity. The Bayes factor for this union is simply the marginal no-bias Bayes factor. It answers whether the data update support toward models without an explicit bias component, but it discards the effect/no-effect resolution that the rigor estimand is designed to preserve.

This distinction is central to the workflow. Component BFs remain useful for diagnosing the evidence profile of an outcome: whether effect evidence, heterogeneity evidence, or explicit bias-component evidence dominates. Rigor names a narrower joint event: resolved effect/no-effect evidence in the branch of the model space that does not require an explicit publication-selection or small-study-effect component.

\subsection{Stratum, corpus, and article-balanced summaries}
\label{sec:rigor-aggregation}

The outcome-level rigor estimand supports summaries under any declared corpus scheme. Let \(S\) denote a scheme and let \(S(k)\) be the stratum assigned to outcome \(k\) under that scheme. Examples include intervention domain, journal window, field-year window, source article, intervention class, outcome family, funder portfolio, or prospective evidence stream. Define the set of outcome indices in stratum \(s\) under scheme \(S\) as
\[
\mathcal{K}_{s,S}=\{k:S(k)=s\},
\qquad
n_{s,S}=|\mathcal{K}_{s,S}|.
\]
The default stratum-level rigor summary is the finite-corpus mean of the selected log rigor Bayes factors:
\[
\Theta_{s,S}^{R}
=
\frac{1}{n_{s,S}}
\sum_{k\in\mathcal{K}_{s,S}}
\log_{10}BF_k^R .
\]
This quantity summarizes the typical log-scale evidential yield among the reconstructed outcome-level evidence objects in stratum \(s\). It is a descriptive property of the analyzed audit corpus. It should not be interpreted as a population estimate unless the corpus was assembled under a sampling design that supports that interpretation.

The same notation also covers corpus-level summaries. If \(s=\mathrm{all}\) denotes the full analyzed corpus under scheme \(S\), then \(\mathcal{K}_{\mathrm{all},S}\) contains all included outcomes and \(\Theta_{\mathrm{all},S}^{R}\) is the corresponding corpus-level mean log rigor. Thus stratum and corpus summaries differ only in which outcome indices are included in \(\mathcal{K}_{s,S}\).

Because the selected branch label is meaningful only together with the rigor magnitude, branch composition is reported separately from the mean log rigor summary. Let
\[
b_k^R
=
\arg\max_{b\in\{+,0\}}
\log_{10}BF_k^{R,b}
\]
denote the selected rigor branch, where \(b_k^R=+\) is the effect branch and \(b_k^R=0\) is the no-effect branch. The unweighted selected-branch proportions are
\[
\pi_{s,S}^{R,+}
=
\frac{1}{n_{s,S}}
\sum_{k\in\mathcal{K}_{s,S}}
\mathbf{1}(b_k^R=+),
\qquad
\pi_{s,S}^{R,0}
=
\frac{1}{n_{s,S}}
\sum_{k\in\mathcal{K}_{s,S}}
\mathbf{1}(b_k^R=0).
\]
These proportions describe which clean branch was selected most often, not whether every selected branch was substantively supported. For that reason, branch composition is interpreted alongside the distribution of \(\log_{10}BF_k^R\) or alongside rigor categories derived from both branch and magnitude. When positive rigor evidence is common, branch composition indicates whether clean evidence accumulation is primarily effect-supporting or no-effect-supporting. When rigor values are near zero or negative, branch labels mainly document which clean branch was less disfavored, rather than providing a substantive effect/no-effect conclusion.

The default summaries above are outcome-weighted: each reconstructed meta-analytic outcome contributes one value of \(\log_{10}BF_k^R\). This is appropriate when the evidence objects of interest are the reconstructed intervention--outcome datasets themselves. However, source articles can contribute unequal numbers of outcomes, so an outcome-weighted average can be influenced by source syntheses with many reconstructed endpoints. To make this design choice explicit, the workflow also reports article-balanced summaries when outcome multiplicity is uneven.

Let \(a(k)\) identify the source meta-analysis article for outcome \(k\), and let \(m_{a(k),s,S}\) be the number of outcomes contributed by that article to stratum \(s\) under scheme \(S\). The default article-balanced weight is
\[
w_k
=
\frac{1}{m_{a(k),s,S}}.
\]
The article-balanced rigor summary is then
\[
\Theta_{s,S}^{R,w}
=
\frac{
	\sum_{k\in\mathcal{K}_{s,S}}
	w_k\log_{10}BF_k^R
}{
	\sum_{k\in\mathcal{K}_{s,S}}w_k
}.
\]
Under this weighting, each source article contributes total weight one within a stratum, regardless of how many reconstructed outcomes it supplied. The corresponding weighted selected-branch proportions are
\[
\pi_{s,S}^{R,+,w}
=
\frac{
	\sum_{k\in\mathcal{K}_{s,S}}
	w_k\mathbf{1}(b_k^R=+)
}{
	\sum_{k\in\mathcal{K}_{s,S}}w_k
},
\qquad
\pi_{s,S}^{R,0,w}
=
\frac{
	\sum_{k\in\mathcal{K}_{s,S}}
	w_k\mathbf{1}(b_k^R=0)
}{
	\sum_{k\in\mathcal{K}_{s,S}}w_k
}.
\]

Outcome-weighted and article-balanced summaries answer related but distinct audit questions. The outcome-weighted summary asks how much clean evidential yield is present among the reconstructed outcome-level evidence objects. The article-balanced summary asks whether the same conclusion is robust after preventing multi-outcome source articles from dominating the stratum. In the present workflow, article-balanced summaries are therefore used as a clustering-aware reporting companion rather than as a replacement for the primary outcome-weighted summaries.

Other weighting systems are possible, including journal-balanced, field-balanced, stratum-balanced, precision-weighted, or user-specified portfolio weights. Those choices change the target of the summary and should be declared before interpretation. The core rigor estimand remains outcome-level; weighting defines how those outcome-level evidence records are summarized for a particular corpus-audit question.

\subsection{Reporting interpretation}
\label{sec:rigor-reporting}

The reporting layer is intentionally simpler than the fitted model space. The sidecars and outcome registry preserve component evidence, joint model-family probabilities, posterior summaries, attenuation measures, branch-specific rigor quantities, diagnostics, and reproducibility metadata. These outputs support secondary analyses and future extensions. The headline reporting quantity, however, is the selected log rigor Bayes factor,
\[
\log_{10}BF_k^R .
\]
Rigor is designed to serve as a common quantitative measure of clean evidential yield: the degree to which an outcome contributes resolved evidence for either an effect or no effect without requiring an explicit publication-selection or small-study-effect component in the fitted ensemble.

Positive values of \(\log_{10}BF_k^R\) indicate movement toward clean resolved evidence. Values near zero indicate little evidential movement toward either clean branch. Negative values indicate that even the better-supported clean branch lost support relative to its complement. Negative rigor should therefore be interpreted as a failure of clean evidential resolution, not as a direct estimate of any single failure mode. Such cases may reflect explicit bias-component support, unresolved effect/no-effect evidence, or other model-space patterns that prevent a clean conclusion.

The selected branch label is retained only to make the scalar rigor value interpretable. When \(\log_{10}BF_k^R\) is positive and substantively large, the branch label identifies whether the clean evidence is effect-supporting or no-effect-supporting. When rigor is near zero or negative, the branch label records which clean branch was less disfavored; it should not be read as a substantive effect or no-effect conclusion. In the machine-readable output contract, this label is stored as \texttt{rigor\_direction}, with values \texttt{effect} and \texttt{no\_effect}.

For compact tables and figures, the workflow also reports a derived rigor category. These categories are display conveniences, not additional estimands. In the present implementation, values satisfying
\[
|\log_{10}BF_k^R|\leq 0.5
\]
are treated as inconclusive with respect to clean evidence, approximately corresponding to Bayes factors between \(1/3\) and \(3\). Values above this region are classified by the selected branch as clean effect-supported or clean no-effect-supported. Values below this region are classified as clean evidence disfavored. The category therefore combines magnitude and branch information in a form suitable for counts, stacked displays, and corpus summaries.

Rigor is interpreted alongside two main diagnostic companions. The first is marginal bias-component evidence, \(\log_{10}BF_k^\omega\), which helps distinguish outcomes whose low rigor is associated with explicit publication-selection or small-study-effect support from outcomes whose low rigor reflects weak or unresolved effect/no-effect evidence. The second is attenuation from the matched Bayesian random-effects baseline to the RoBMA-PSMA estimate. Attenuation is especially important for apparent positive effects: it shows whether an unadjusted effect survives bias-aware fitting or is reduced under the bias-aware ensemble. Attenuation is not the primary evidential-yield measure, but it is a key explanatory companion for effect-branch cases.

At stratum and corpus scales, the main object of comparison is the distribution or summary of \(\log_{10}BF_k^R\). This is what allows the workflow to compare evidence accumulation across intervention classes, journals, field-year windows, research programs, funder portfolios, or prospective evidence streams. Before bias-aware model-family evidence was available, such comparisons were largely qualitative: analysts could inspect significance patterns, risk-of-bias judgments, funnel-plot diagnostics, and narrative certainty assessments, but there was no single model-based evidence quantity that jointly represented resolved effect/no-effect evidence and absence of an explicit bias component. Rigor supplies that common quantitative target.

The simulation and resampling layer evaluates whether these summaries are stable enough to support such comparisons. Its central stability measure is interval width for the rigor summary, especially
\[
\mathrm{width}_{90}=q_{0.95}-q_{0.05}.
\]
Rigor width is not a competing estimand. It is a performance measure for the workflow: it describes how variable stratum- or corpus-level rigor summaries are under known-cell synthetic sampling, empirical outcome or source-cluster resampling, and empirical fitted-profile-weighted synthetic sampling. This is particularly important in the small-to-moderate \(n_{\mathrm{outcomes}}\) regimes expected in practical evidential audits.

The resulting hierarchy is deliberate. Rigor is the headline measure of effective evidence accumulation. Branch labels and categories make rigor readable. Bias-component evidence and attenuation explain why rigor is high or low. Resampling width evaluates whether summaries of rigor are stable enough for comparison. The richer sidecar and registry outputs remain available for diagnostic work, sensitivity analysis, and future extensions, but the core reporting target is intentionally compact.

\section{Simulation and resampling characterization}
\label{sec:simulation}

\subsection{Purpose and design stance}
\label{sec:simulation-purpose}

We use the ADEMP framework of \citet{Morris2019} to structure the simulation and resampling characterization. The framework begins with aims: what the simulation study is intended to learn before specifying data-generating or resampling mechanisms, estimands, methods, and performance measures. In this paper, Q1--Q3 are the three planned aims of the evaluation layer. They are not merely figure groups or script outputs; they define the questions that the simulation/resampling layer is designed to answer.

The object being evaluated is the assembled evidential-audit workflow rather than a newly proposed estimator in isolation. Across all three aims, the same method is held fixed: audit-ready study-level effect sizes and standard errors are passed through the matched Bayesian random-effects baseline, the RoBMA-PSMA ensemble, sidecar reduction, registry rebuild, rigor-first reducers, and generated reporting layer. What changes across Q1--Q3 is the mechanism used to generate or resample outcome-level evidence objects.

\paragraph{Q1. Known-cell operating behavior.}
\emph{Question:} Under known synthetic effect, heterogeneity, and bias-burden regimes, do the matched baseline and RoBMA-PSMA audit outputs behave as expected?  

\emph{Aim:} Q1 provides a local operating check for the workflow under the synthetic DGM used in this paper. It is not intended to revalidate RoBMA-PSMA from first principles; the RoBMA literature already provides a much broader evaluation of the model family. Instead, Q1 asks whether, under our 36-cell design, the matched random-effects baseline and the bias-aware RoBMA-PSMA fit recover and respond to the known generating structure in ways that make the downstream audit quantities interpretable. This aim grounds the attenuation atlas, the effect-recovery comparison between baseline and bias-aware fits, and the known-cell selected-rigor displays.

\paragraph{Q2. Empirical-registry stability.}
\emph{Question:} In the observed nutrition registry, how stable are selected rigor and its main explanatory companions under realistic finite-registry resampling?  

\emph{Aim:} Q2 treats the empirical nutrition corpus as the relevant finite evidence registry and studies how the reported stratum and corpus summaries behave under outcome-row and source-article cluster resampling. Because the empirical data-generating mechanism is unknown, Q2 is not a known-truth simulation and does not convert the purposive nutrition corpus into a population sample. Its purpose is to assess whether selected rigor, bias evidence, and attenuation are stable features of the analyzed registry, and whether summaries are sensitive to outcome multiplicity or influential source syntheses.

\paragraph{Q3. Empirical fitted-profile-weighted synthetic sampling.}
\emph{Question:} Beyond the exact empirical registry, how do selected rigor and companion audit summaries behave when synthetic outcomes are resampled under empirical fitted-profile structure?  

\emph{Aim:} Q3 extends the characterization beyond the observed nutrition outcomes by projecting empirical outcomes into fitted-profile cells and using those fitted-profile mixtures to sample from the known-cell synthetic library. This is not a new DGM and not a calibration claim that empirical outcomes came from the synthetic cells. Instead, it asks whether the rigor workflow remains stable and interpretable when applied to synthetic outcome draws with empirical-like fitted-profile structure. Q3 therefore connects the empirical case study to a wider set of synthetic evidence structures and provides a broader check on the portability of the rigor estimand and its companion summaries.

\begin{figure}[H]
	\centering
	\includegraphics[width=\linewidth,height=1\textheight,keepaspectratio]{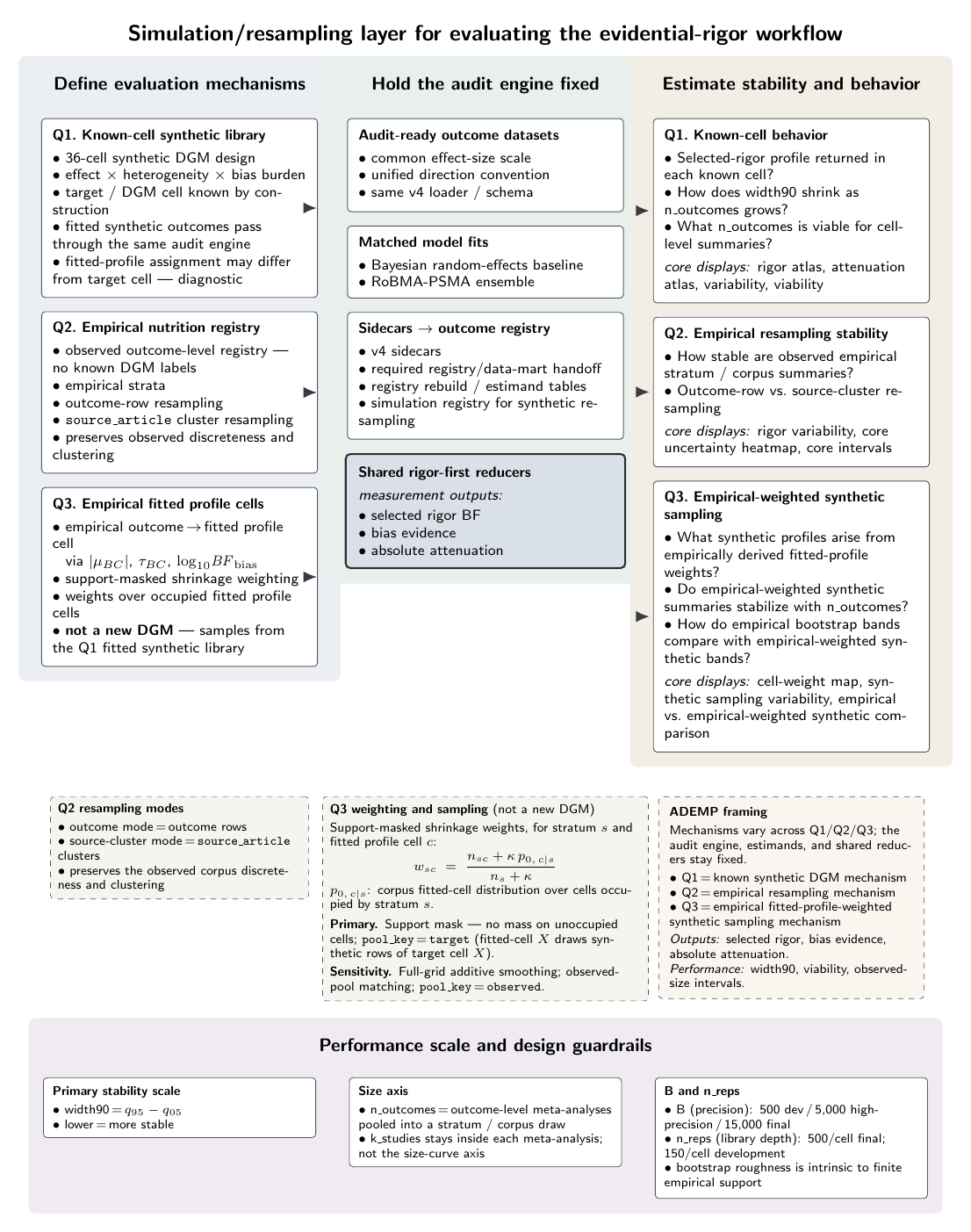}
	\caption{Simulation/resampling evaluation architecture. The three evaluation mechanisms differ in how outcome-level evidence objects are generated or resampled, while the audit engine, estimands, and reducers are held fixed.}
	\label{fig:simulation-resampling-architecture}
\end{figure}

Figure~\ref{fig:simulation-resampling-architecture} summarizes this design. The left column separates the three evaluation mechanisms, the middle column emphasizes that the audit engine and reducers are fixed, and the right column links each aim to the behavior and stability summaries reported below. The bottom panels record the resampling modes, Q3 weighting rule, ADEMP mapping, and the distinction between \(B\), which controls resampling or sampling precision, and \(n_{\mathrm{reps}}\), which controls fitted synthetic-library depth.

This structure follows the logic of ADEMP. The aims determine why Q1--Q3 are needed; the data-generating and resampling mechanisms determine how evidence objects are produced; the estimands determine what is summarized; the method is the fixed audit workflow; and the performance measures determine how behavior and stability are judged.

\subsection{Data-generating and resampling mechanisms}
\label{sec:simulation-aims}
\label{sec:simulation-mechanisms}
\label{sec:fitted-profile-cells}
\label{sec:fitted-cell-weights}

After specifying the three aims, the next ADEMP element is the mechanism used to generate or resample the evidence objects. The mechanisms differ across Q1--Q3 because the aims differ. Q1 uses a known synthetic DGM to study workflow behavior under controlled truth. Q2 uses empirical resampling because the real-world data-generating mechanism for the nutrition registry is unknown. Q3 uses empirical fitted-profile-weighted synthetic sampling to connect empirical registry structure to the broader known-cell synthetic library.

\paragraph{Q1 mechanism: known synthetic DGM cells.}
For Q1, each generated dataset belongs to a known target cell. The synthetic design crosses four effect levels, three heterogeneity levels, and three bias-burden levels, yielding \(4 \times 3 \times 3 = 36\) target cells. The effect anchors are
\[
\mu_{\mathrm{true}} \in \{0.00, 0.10, 0.25, 0.50\},
\]
the heterogeneity anchors are
\[
\tau_{\mathrm{true}} \in \{0.05, 0.15, 0.40\},
\]
and the bias-burden axis combines threshold-selection structure with small-study-effect strength. Each generated audit-ready dataset is then passed through the same loader, matched Bayesian random-effects baseline, RoBMA-PSMA ensemble, sidecar writer, registry builder, and summary reducers used elsewhere in the workflow. Thus the simulator knows the target cell, but the fitted audit workflow sees only the study-level effect sizes and standard errors.

\paragraph{Q2 mechanism: empirical registry resampling.}
For Q2, the data-generating mechanism is not known. The evidence objects are the reconstructed outcome-level meta-analyses in the empirical nutrition registry. We therefore study finite-registry stability by resampling the observed registry rather than by assigning empirical outcomes to a known truth cell. Two resampling modes are used. In the outcome-row mode, outcome-level meta-analyses are resampled with replacement within stratum. In the source-cluster mode, \texttt{source\_article} clusters are resampled with replacement within stratum. The source-cluster mode preserves article-level outcome multiplicity and identifies summaries that are sensitive to influential source syntheses.

\paragraph{Q3 mechanism: empirical fitted-profile-weighted synthetic sampling.}
For Q3, empirical outcomes are first projected into fitted-profile cells using fitted audit outputs. Stratum-specific fitted-profile mixtures are then used to sample fitted synthetic outcomes from the Q1 library. The primary pool basis is the known synthetic target cell, so Q3 samples from synthetic outcomes with known generating structure while weighting those samples by empirical fitted-profile structure. This mechanism is intentionally intermediate between Q1 and Q2: it is synthetic and known-cell at the pool level, but empirical-profile-weighted at the stratum level.

The fitted-profile projection uses the same 36-cell vocabulary as the synthetic DGM, but the meanings differ. For synthetic datasets, the target cell is a known DGM label. For empirical outcomes, the fitted-profile cell is a deterministic coarsening of fitted audit outputs. It is not a latent truth label and should not be interpreted as evidence that the empirical outcome was generated by the corresponding synthetic mechanism.

For empirical outcome \(k\), define the fitted-profile cell
\[
c_k^{\mathrm{fit}}
=
\left(e_k^{\mathrm{fit}},h_k^{\mathrm{fit}},a_k^{\mathrm{fit}}\right),
\]
where \(e_k^{\mathrm{fit}}\) is the fitted effect band, \(h_k^{\mathrm{fit}}\) is the fitted heterogeneity band, and \(a_k^{\mathrm{fit}}\) is the fitted bias-evidence band. These bands are defined by
\[
e_k^{\mathrm{fit}}
=
\begin{cases}
	\text{null}, & |\mu_{\mathrm{BC},k}| < 0.05,\\
	\text{small}, & 0.05 \le |\mu_{\mathrm{BC},k}| < 0.175,\\
	\text{moderate}, & 0.175 \le |\mu_{\mathrm{BC},k}| < 0.375,\\
	\text{large}, & |\mu_{\mathrm{BC},k}| \ge 0.375,
\end{cases}
\]
\[
h_k^{\mathrm{fit}}
=
\begin{cases}
	\text{lowhet}, & \tau_{\mathrm{BC},k} < 0.10,\\
	\text{midhet}, & 0.10 \le \tau_{\mathrm{BC},k} < 0.275,\\
	\text{highhet}, & \tau_{\mathrm{BC},k} \ge 0.275,
\end{cases}
\]
and
\[
a_k^{\mathrm{fit}}
=
\begin{cases}
	\text{clean}, & \log_{10}BF_{\mathrm{bias},k} \le 0.5,\\
	\text{modbias}, & 0.5 < \log_{10}BF_{\mathrm{bias},k} \le 1.0,\\
	\text{highbias}, & \log_{10}BF_{\mathrm{bias},k} > 1.0 \text{ or } \log_{10}BF_{\mathrm{bias},k}=\infty .
\end{cases}
\]
These bands define a fitted-profile index, not a substantive classification of study validity. In particular, the clean band means that the fitted model-family evidence does not favor an explicit bias component under the RoBMA-PSMA ensemble; it does not prove absence of all possible bias mechanisms.

For Q3, the primary stratum-specific sampling weights use support-masked shrinkage weighting (recorded in code as the empirical-Bayes corpus rule \texttt{eb\_corpus}). Let \(n_{sc}\) be the number of empirical outcomes in stratum \(s\) assigned to fitted-profile cell \(c\), let \(n_s\) be the total number of empirical outcomes in stratum \(s\), and let \(p_{0,c\mid s}\) denote the corpus fitted-cell distribution restricted and renormalized over cells occupied by stratum \(s\). The Q3 sampling weight is
\[
w_{sc}
=
\frac{n_{sc} + \kappa p_{0,c\mid s}}
{n_s + \kappa}.
\]
Here \(\kappa=4\) is the corpus-prior strength measured in pseudo-outcomes. The shrinkage is applied to the empirical fitted-profile cell counts within each stratum, so the effective empirical weight is \(n_s/(n_s+\kappa)\) and smaller strata shrink more strongly toward the restricted corpus distribution. The support mask restricts smoothing to cells already occupied by that stratum, so the weighting rule stabilizes sparse empirical mixtures without inventing mass in unobserved fitted-profile regions. The resulting weights are then used to sample fitted synthetic outcomes from the Q1 library; they are not estimates of latent empirical DGM membership. Full-grid additive smoothing and matching on observed fitted synthetic cells rather than target cells are retained as sensitivity analyses, not as the primary specification. The primary Q3 setting is \texttt{pool\_key = target}, \texttt{smoothing = eb\_corpus}, \(\kappa=4\), and \texttt{support = occupied}.

Throughout these mechanisms, the relevant size axis for stability is \(n_{\mathrm{outcomes}}\), the number of outcome-level meta-analyses summarized inside a stratum or corpus draw. This is distinct from \(k_{\mathrm{studies}}\), the number of study-level estimates inside one meta-analytic outcome. The synthetic DGM draws \(k_{\mathrm{studies}}\) inside each generated outcome; the simulation and resampling characterization primarily studies how stratum-level summaries behave as \(n_{\mathrm{outcomes}}\) varies.

\subsection{Estimands and evaluated method}
\label{sec:simulation-estimands}

The estimand layer is fixed across Q1--Q3. The primary estimand propagated through the simulation and resampling characterization is selected rigor, \(\log_{10}BF_k^R\), as defined in Section~\ref{sec:rigor-estimand}. The selected branch label \(b_k^R\) is retained as interpretive metadata, but the headline quantity is the selected \(\log_{10}\) rigor Bayes factor. The branch label is substantively interpretable only alongside the sign and magnitude of \(\log_{10}BF_k^R\). A negative value does not indicate evidence for no effect; it indicates that clean resolved evidence, under either branch, is disfavored.

The main outcome-level quantities carried into the simulation and resampling summaries are selected rigor evidence, selected branch label, bias-component evidence, and absolute attenuation between the matched random-effects baseline and the RoBMA-PSMA estimate. Bias evidence and attenuation are explanatory companions. They help interpret why apparent effects may lose clean evidential support, but they are not the headline evidential-yield estimands. Rigor margin is retained as a sidecar diagnostic field in the outcome registry but is not a manuscript-facing core-metric panel.

The evaluated method is also fixed across Q1--Q3. Each outcome-level evidence object is represented by audit-ready study-level effect sizes and standard errors; fitted using the matched Bayesian random-effects baseline and RoBMA-PSMA ensemble; reduced to a sidecar; rebuilt into the canonical registry; and summarized using the same rigor-first reducers. Thus the simulation/resampling layer evaluates the assembled workflow as used in the empirical case study, not a separate simulation-only implementation.

\subsection{Performance measures, computational scale, and reporting}
\label{sec:simulation-repetition}
\label{sec:simulation-reporting}

The primary stability measure is
\[
\mathrm{width90} = q_{0.95} - q_{0.05},
\]
computed across resampling or sampling draws for a specified stratum, corpus, metric, and sample size. Lower \(\mathrm{width90}\) indicates a more stable summary. This is a resampling or sampling variability measure across \(B\) draws; it is not posterior uncertainty and it is not itself a Monte Carlo standard error.

Additional summaries include observed-size intervals for empirical resampling, minimum-\(n_{\mathrm{outcomes}}\) viability summaries for known-cell behavior, and empirical-versus-synthetic agreement summaries for Q3 bridge checks. Binary rates are retained where useful, but the main figures emphasize continuous rigor evidence and continuous core audit metrics rather than thresholded positive-finding rates. For binary rate summaries, Monte Carlo uncertainty is monitored using the binomial Monte Carlo standard error,
\[
\mathrm{MCSE}(\hat p)=\sqrt{\frac{\hat p(1-\hat p)}{B}},
\]
where \(B\) is the number of resampling or sampling draws. The worst-case binomial MCSE occurs at \(\hat p=0.5\) and equals \(\sqrt{0.25/B}\). At the publication setting \(B=\SIMB\) the worst-case binary MCSE is therefore approximately \(\SIMMCSEfinal\); the internal high-precision tier (\(B=\SIMBcheck\)) gives \(\SIMMCSEcheck\), and the development/visual-tuning tier (\(B=\SIMBdev\)) gives \(\SIMMCSEdev\). For continuous summaries, \(B\) controls the precision of the empirical quantiles used to compute \(\mathrm{width90}\); the displayed \(\mathrm{width90}\) is the resampling/sampling variability defined above, not a Monte Carlo standard error.

Two repetition quantities are separated. The fitted-library depth \(n_{\mathrm{reps}}\) controls how many synthetic outcomes are generated and fit within each Q1 target cell. The resampling count \(B\) controls MC precision for Q2 empirical resampling and Q3 empirical fitted-profile-weighted synthetic sampling. The publication configuration uses \(n_{\mathrm{reps}}=\SIMnreps\) fitted synthetic outcomes per cell, yielding \SIMnSyntheticOutcomes{} fitted synthetic outcomes across the \SIMnCells-cell design, and \(B=\SIMB\) resampling or sampling draws for the final Q2 and Q3 summaries. Development and internal-check runs use \(n_{\mathrm{reps}}=\SIMnrepsdev\), \(B=\SIMBdev\), or \(B=\SIMBcheck\) for visual tuning and diagnostic iteration. Figure annotations record the actual analysis tier used to generate the displayed artifact; the centralized definitions above govern the interpretation of \(\mathrm{width90}\) and the binary MCSE in every Section~4 figure.

Simulation and resampling results are reported using generated figures and CSV-backed summaries. Each figure records the relevant design metadata in its annotation: the resampling count \(B\), the fitted-library depth or available pool size where relevant, the Q2 resampling mode, the Q3 pool basis, smoothing rule, \(\kappa\), support restriction, and source CSV. This convention is part of the reproducibility design. The \LaTeX{} manuscript references stable artifact basenames, while the repository records the producer scripts, source CSVs, and sync manifest used to generate and pin each figure.

The main Q1 figures summarize known-cell selected-rigor behavior, companion attenuation/effect recovery, and stability over \(n_{\mathrm{outcomes}}\). The main Q2 figures summarize empirical resampling variability under outcome-row and source-cluster resampling. The main Q3 figures summarize empirical fitted-cell mixture weights, empirical-weighted synthetic stability, and bridge comparisons between empirical bootstrap summaries and empirical-weighted synthetic summaries. Supplementary diagnostics report fitted-profile recovery, pool support, and additional component-stability summaries.

\subsection{Q1: known-cell operating behavior}
\label{sec:q1-known-cell-behavior}

Q1 evaluates the full audit workflow under known synthetic target cells. This part of the characterization has a deliberately limited role. It is not a new large-scale validation of RoBMA-PSMA and is not meant to replace the broader simulation literature for the model family. Instead, it asks whether the synthetic cells used in this evaluation produce interpretable behavior from the assembled workflow: the matched random-effects baseline, the RoBMA-PSMA ensemble, the sidecar and registry reductions, and the selected-rigor summaries.

\begin{figure}[H]
	\centering
	\includegraphics[width=\linewidth]{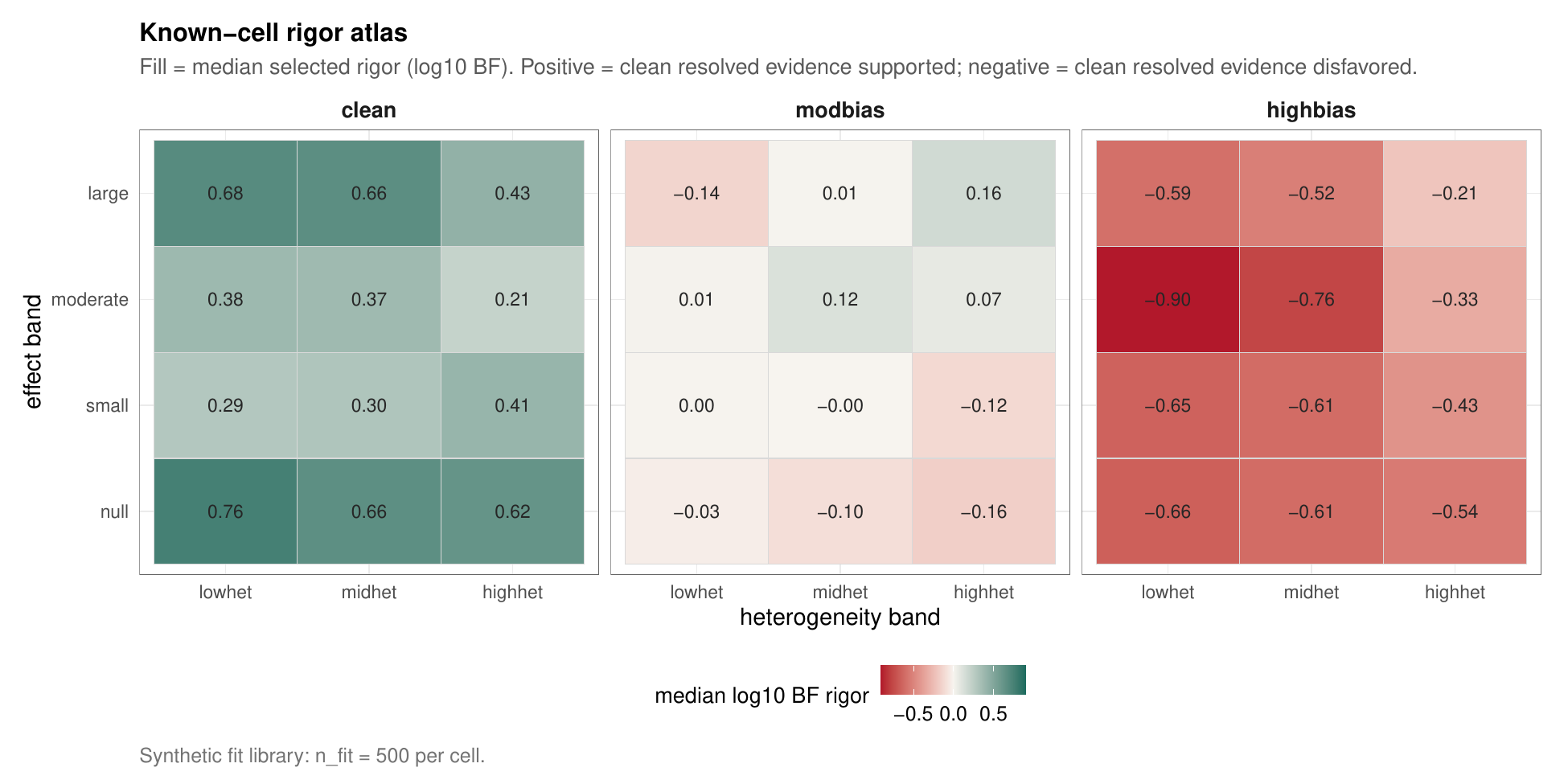}
	\caption{Q1 known-cell selected-rigor atlas. Each panel summarizes \(\log_{10}BF_k^R\) for fitted synthetic outcomes generated under a known effect \(\times\) heterogeneity \(\times\) bias-burden target cell. The figure evaluates whether the workflow returns coherent selected-rigor evidence across the 36-cell design.}
	\label{fig:q1-rigor-atlas}
\end{figure}

The first display is the selected-rigor atlas in Figure~\ref{fig:q1-rigor-atlas}. The main pattern is that selected rigor is easiest to resolve when the generating structure supplies enough information for a clean effect or clean no-effect branch to dominate. Small-effect cells are correspondingly more difficult: even under relatively favorable conditions, weak effects can leave the fitted ensemble without strong resolved evidence, and selected rigor remains limited. This is an important feature rather than a defect. Rigor is not an effect-size estimator and not a positive-finding score; it is a joint evidence measure that requires effect/no-effect resolution together with absence of an explicit bias component. Clean generating conditions therefore do not automatically imply high rigor. Clean but weakly informative evidence can still be evidentially unresolved.

The bias-burden axis behaves more predictably. Cells with stronger bias structure tend to produce lower clean resolved support, stronger bias-component evidence, or both. Thus, the atlas does two things at once: it shows that the workflow can recover coherent clean evidence where the known cell supplies enough information, and it shows that the rigor estimand appropriately withholds clean evidential credit when the generating structure makes bias-aware resolution difficult.

\begin{figure}[H]
	\centering
	\includegraphics[width=\linewidth]{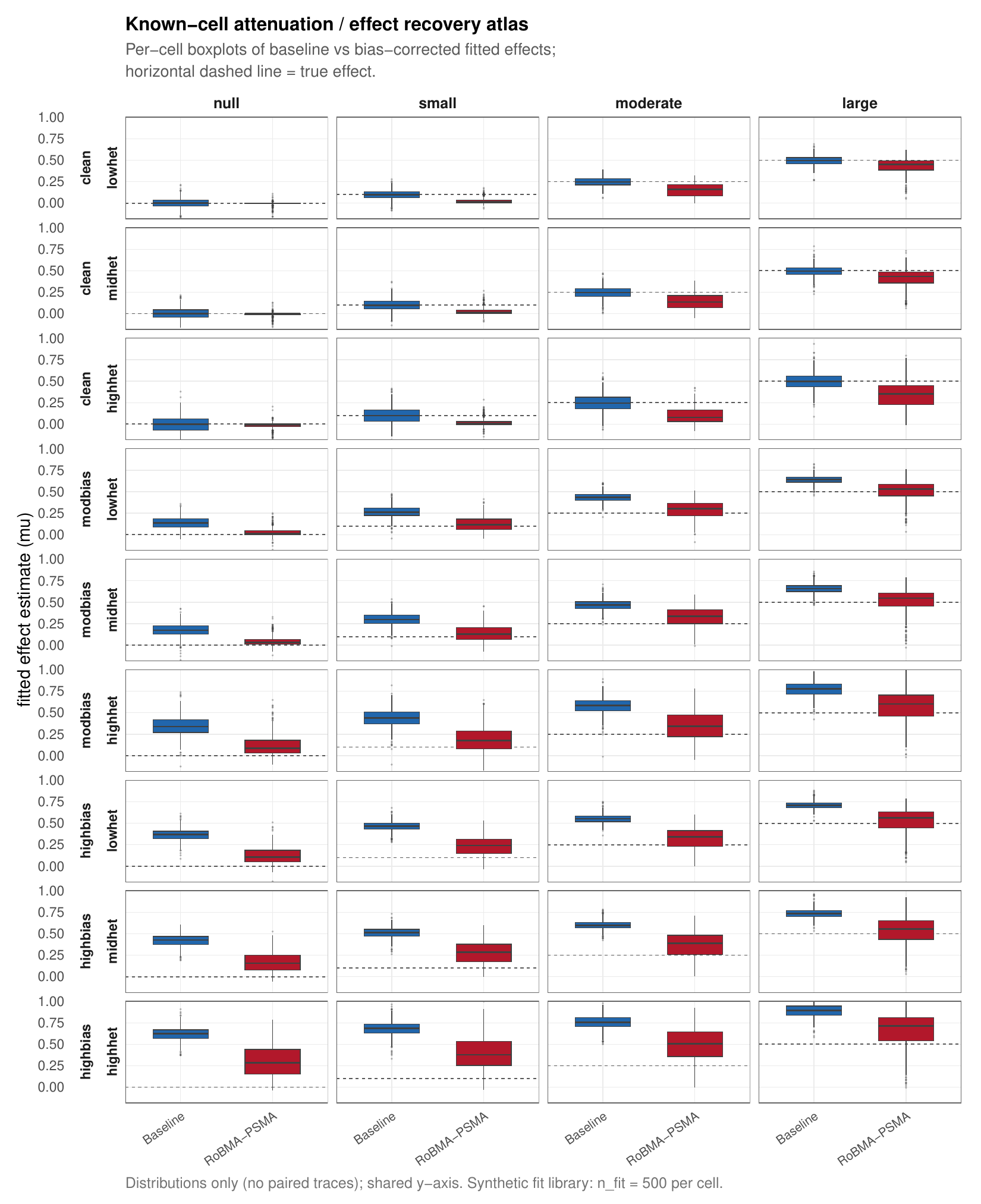}
	\caption{Q1 companion attenuation and effect-recovery atlas. The figure compares the matched Bayesian random-effects baseline with the RoBMA-PSMA estimate across the 36 known synthetic cells. Attenuation is interpreted as an explanatory audit companion to selected rigor, not as the primary evidential-yield estimand.}
	\label{fig:q1-attenuation-atlas}
\end{figure}

Figure~\ref{fig:q1-attenuation-atlas} gives the companion effect-recovery and attenuation display.It shows whether the matched random-effects baseline and the bias-aware RoBMA-PSMA fit respond sensibly to the generating structure. Under relatively clean conditions, the two fitted summaries should be broadly consistent with the target structure. Under stronger bias burden, divergence between the baseline and the bias-aware fit is expected rather than problematic: the bias-aware model should reduce apparent effects when the visible study record is generated with bias-relevant structure.

The attenuation atlas therefore grounds one of the main interpretive uses of the empirical case study. In Section~\ref{sec:case-study}, attenuation is used to explain why nominally meaningful baseline effects may lose clean evidential support after bias-aware fitting. Q1 shows this behavior under known synthetic conditions before the workflow is applied to the empirical registry. At the same time, attenuation remains an explanatory companion. It does not replace selected rigor as the evidential-yield estimand, especially in settings where successful evidence accumulation may consist of ruling out an effect.

\begin{figure}[H]
	\centering
	\includegraphics[width=\linewidth]{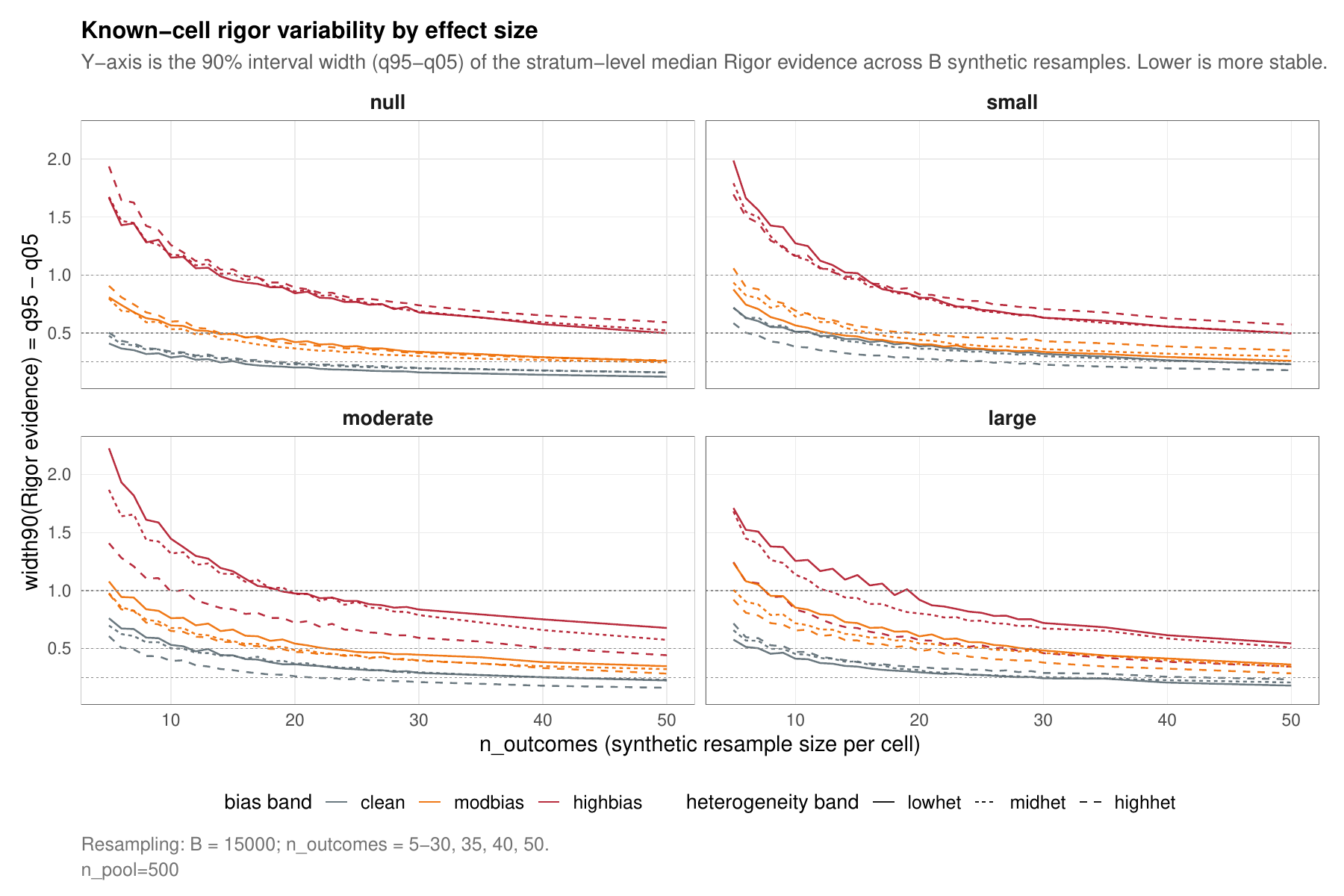}
	\caption{Q1 selected-rigor variability over \(n_{\mathrm{outcomes}}\). The vertical scale is \(\mathrm{width90}=q_{0.95}-q_{0.05}\) for stratum-level selected rigor across repeated synthetic draws. Lower values indicate more stable summaries.}
	\label{fig:q1-rigor-variability}
\end{figure}

The next Q1 question is stability: how much does a stratum-level selected-rigor summary vary as the number of outcome-level meta-analyses increases? Figure~\ref{fig:q1-rigor-variability} reports \(\mathrm{width90}\) for selected rigor across the synthetic size curve. The relevant size axis is \(n_{\mathrm{outcomes}}\), not the number of studies inside a single meta-analysis. As \(n_{\mathrm{outcomes}}\) grows, the sampling variability of the stratum-level selected-rigor summary generally contracts. The figure therefore provides a scale for interpretation: apparent differences in stratum-level rigor should be interpreted relative to the stability one would expect at comparable outcome counts and fitted-profile structure.

This display does not create a universal threshold for meaningful differences. A difference of one-half log unit in selected rigor may be substantial in a stable region of the design and less compelling in a region where the width curve remains broad. The point of Figure~\ref{fig:q1-rigor-variability} is to make that dependence visible. It shows where selected-rigor summaries stabilize quickly and where weak effects, heterogeneity, or bias burden make stratum-level summaries more variable.

\begin{figure}[H]
	\centering
	\includegraphics[width=\linewidth]{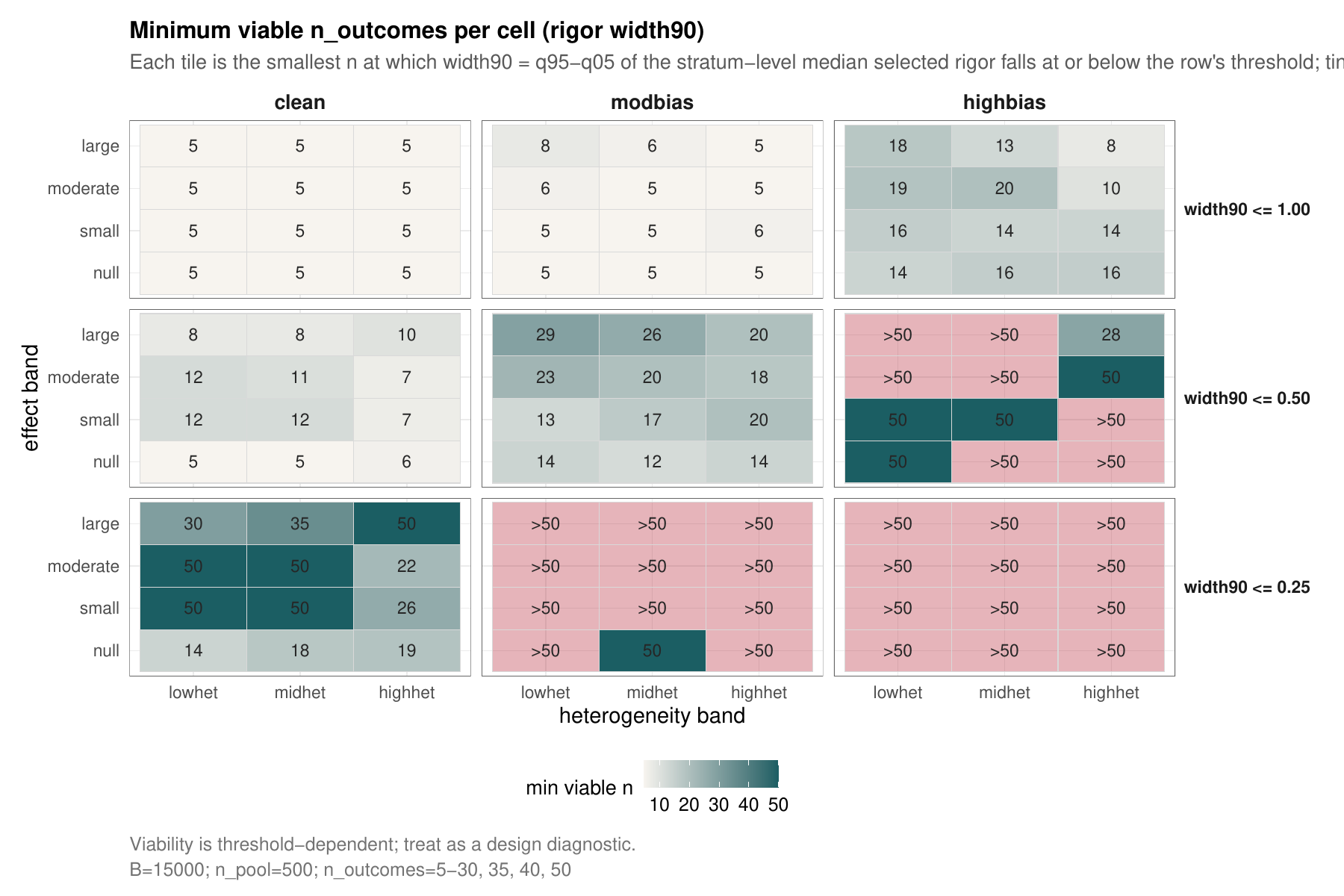}
	\caption{Q1 minimum-\(n_{\mathrm{outcomes}}\) viability summary for selected rigor. The figure reports the smallest outcome count at which the selected-rigor summary meets the stated stability criterion within each known synthetic cell, with cells that do not meet the criterion inside the evaluated grid marked accordingly.}
	\label{fig:q1-rigor-viability}
\end{figure}

Figure~\ref{fig:q1-rigor-viability} turns the same stability question into a minimum-\(n_{\mathrm{outcomes}}\) display. Rather than asking only whether the width curve eventually decreases, the viability plot asks how many outcome-level meta-analyses are needed before the selected-rigor summary meets the stated width criterion. This is one of the most practically useful simulation outputs for corpus-scale auditing. Many real audit strata will be small to moderate in size; the viability display makes explicit which known-cell structures can support stable stratum-level rigor summaries at those sizes and which require more evidence.

Taken together, the Q1 figures provide a local operating check for the workflow. They show that selected rigor behaves coherently under known target cells, that the baseline and bias-aware estimates respond sensibly to bias burden, and that the stability of stratum-level rigor summaries depends on both outcome count and generating structure. The main caution is that weak effects remain hard to resolve. This is exactly the setting in which a scalar evidential-yield measure is useful: it prevents small, unstable, or bias-dependent apparent effects from being counted as clean evidence merely because an effect estimate is nonzero.

\subsection{Q2: empirical-registry stability}
\label{sec:q2-empirical-resampling}

Q2 moves from known synthetic cells to the observed nutrition registry. Here the mechanism is empirical resampling, and the target is finite-registry stability: how much do observed stratum- and corpus-level summaries vary under plausible resampling views of the analyzed registry? Outcome-row resampling treats each reconstructed outcome as the resampling unit, while source-cluster resampling treats source articles as the resampling unit. Comparing the two modes distinguishes ordinary finite-outcome variability from dependence on particular source syntheses.

\begin{figure}[H]
	\centering
	\includegraphics[width=\linewidth]{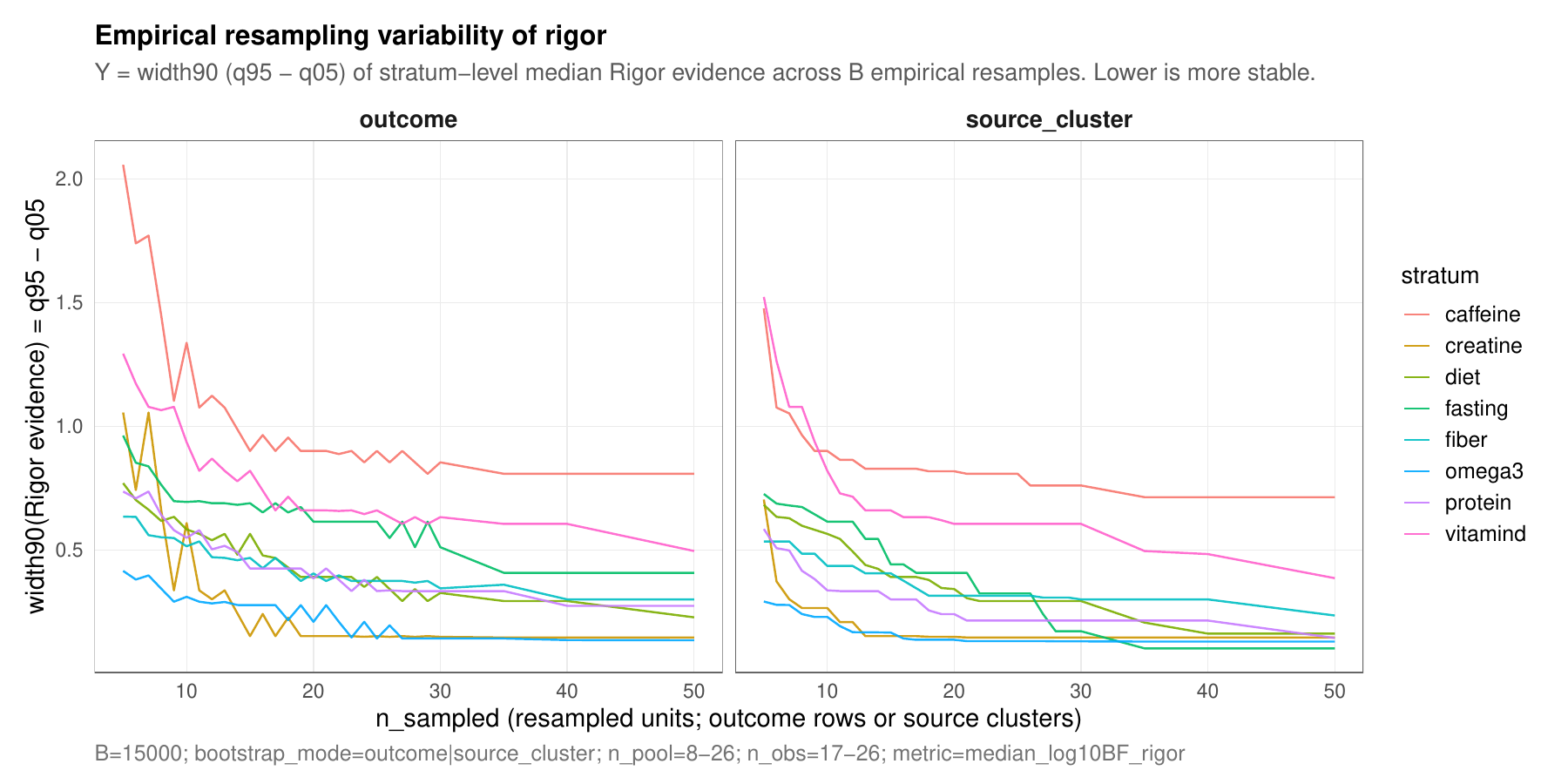}
	\caption{Q2 empirical resampling variability for selected rigor by stratum. Outcome-row and source-cluster resampling estimate how stable observed nutrition stratum summaries are under finite-registry resampling.}
	\label{fig:q2-rigor-variability}
\end{figure}

Figure~\ref{fig:q2-rigor-variability} is the empirical counterpart to the Q1 size-curve display. The figure uses the same \(\mathrm{width90}\) scale, but now the variation comes from resampling the observed nutrition registry rather than from drawing synthetic outcomes from a known cell. This is the first check on whether the empirical case-study summaries are stable enough to interpret. Strata with narrower curves are less sensitive to the finite composition of the registry; strata with wider curves require more caution, especially if outcome-row and source-cluster resampling diverge.

The source-cluster comparison is particularly important. A stratum can look stable under outcome-row resampling while still depending heavily on one or two multi-outcome source syntheses. When source-cluster resampling is wider, the issue is not merely small \(n_{\mathrm{outcomes}}\); it is article-level dependence. This distinction matters for the case study because the nutrition registry is purposive and finite, not a probability sample of all nutrition evidence.

\begin{figure}[H]
	\centering
	\includegraphics[width=\linewidth]{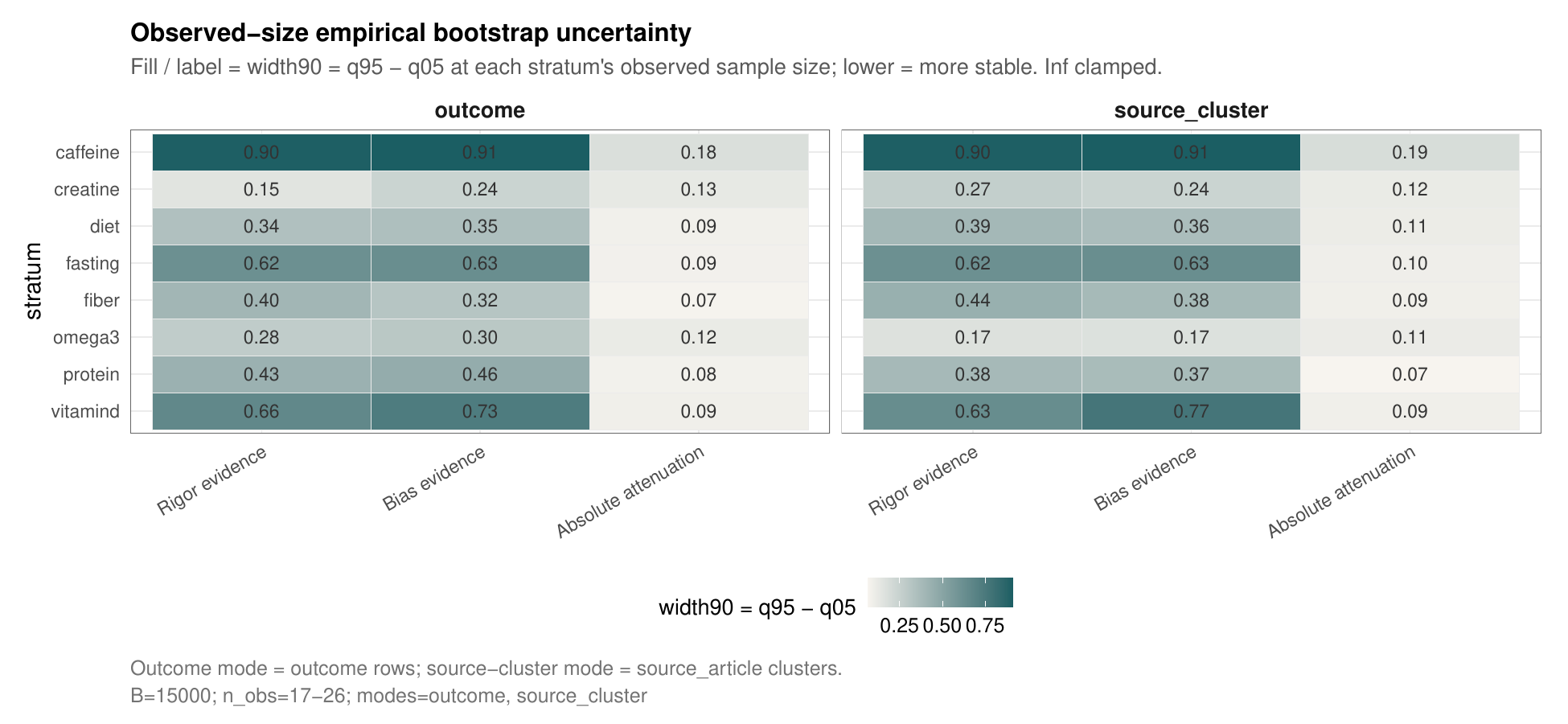}
	\caption{Q2 observed-size uncertainty heatmap for selected rigor and diagnostic companions. The display summarizes empirical resampling \(\mathrm{width90}\) for rigor evidence, bias evidence, and absolute attenuation across strata at the observed stratum sizes.}
	\label{fig:q2-core-uncertainty}
\end{figure}

\begin{figure}[H]
	\centering
	\includegraphics[width=\linewidth]{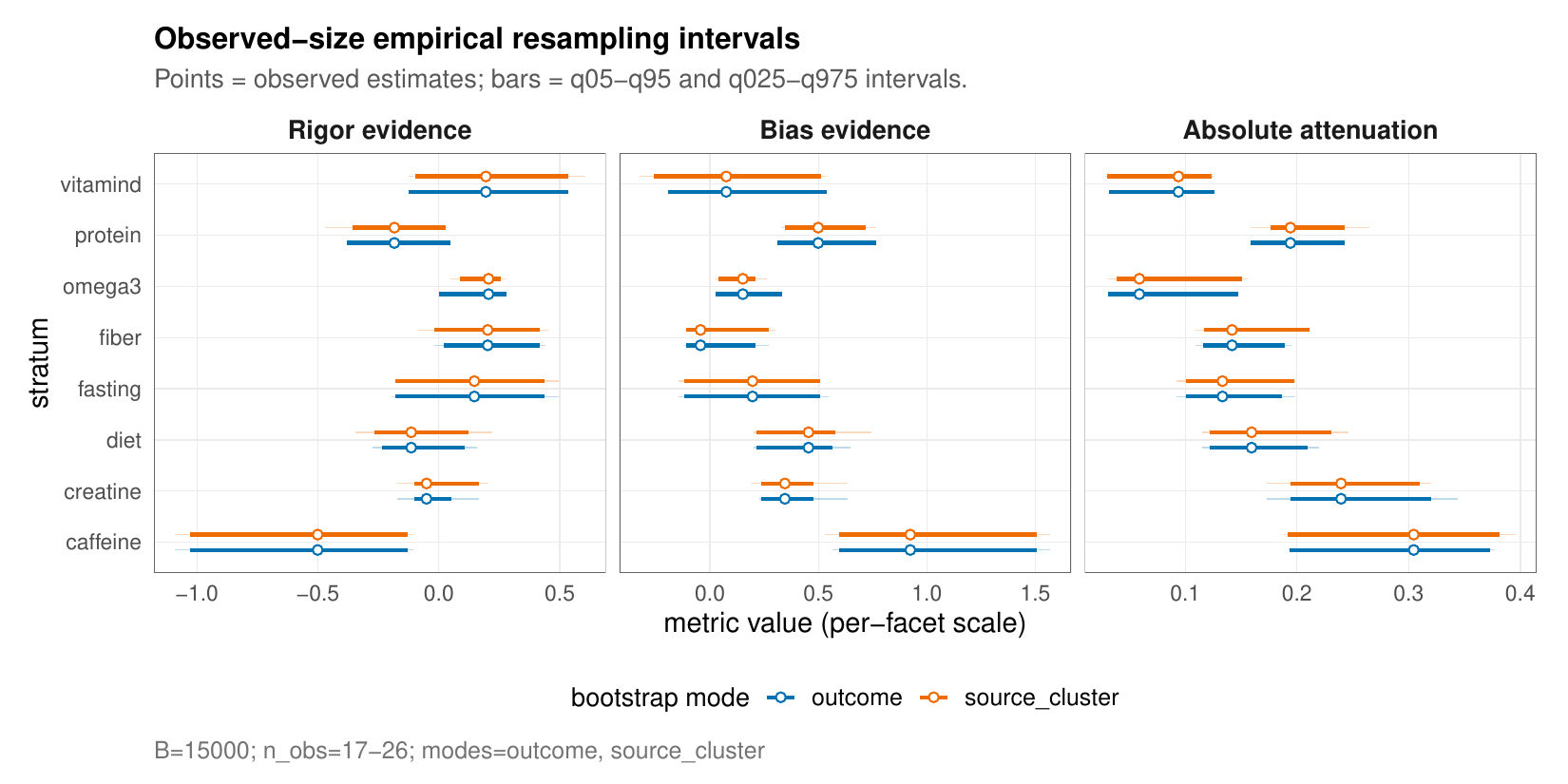}
	\caption{Q2 observed-size empirical resampling intervals for selected rigor and diagnostic companions. The figure reports finite-registry uncertainty for rigor evidence, bias evidence, and absolute attenuation under the empirical resampling design.}
	\label{fig:q2-core-intervals}
\end{figure}

Figures~\ref{fig:q2-core-uncertainty} and~\ref{fig:q2-core-intervals} extend the empirical stability check to selected rigor and two diagnostic companions. The load-bearing panel is selected rigor, because it is the evidential-yield estimand carried forward into the case study. Bias evidence is the main model-family companion: it helps distinguish low rigor associated with explicit publication-selection or small-study-effect support from low rigor associated with weak or unresolved effect/no-effect evidence. Absolute attenuation is an effect-movement companion: it describes how much the matched random-effects baseline changes under bias-aware fitting.

These companion panels should be interpreted diagnostically rather than as competing evidence measures. Bias evidence is useful because it directly explains one pathway through which clean resolved evidence can fail. Absolute attenuation is useful because it identifies settings where baseline effects are reduced after bias-aware fitting, but attenuation is not intrinsically good or bad and should not be treated as a general measure of evidential success. Rigor margin is retained in the outcome registry as a sidecar diagnostic, but it is not used as a manuscript-facing core-metric panel.

The Q2 results are not intended to convert the purposive nutrition corpus into a population sample of nutrition evidence. They instead quantify how stable the reported empirical summaries are under two reasonable resampling views of the observed registry. This is the practical empirical question that must be answered before using the case study descriptively: are the observed stratum summaries stable enough to discuss, and where do source clustering or finite registry size make the summary fragile?

\subsection{Q3: empirical fitted-profile-weighted synthetic sampling}
\label{sec:q3-empirical-weighted-synthetic}

\begin{figure}[H]
	\centering
	\includegraphics[width=\linewidth]{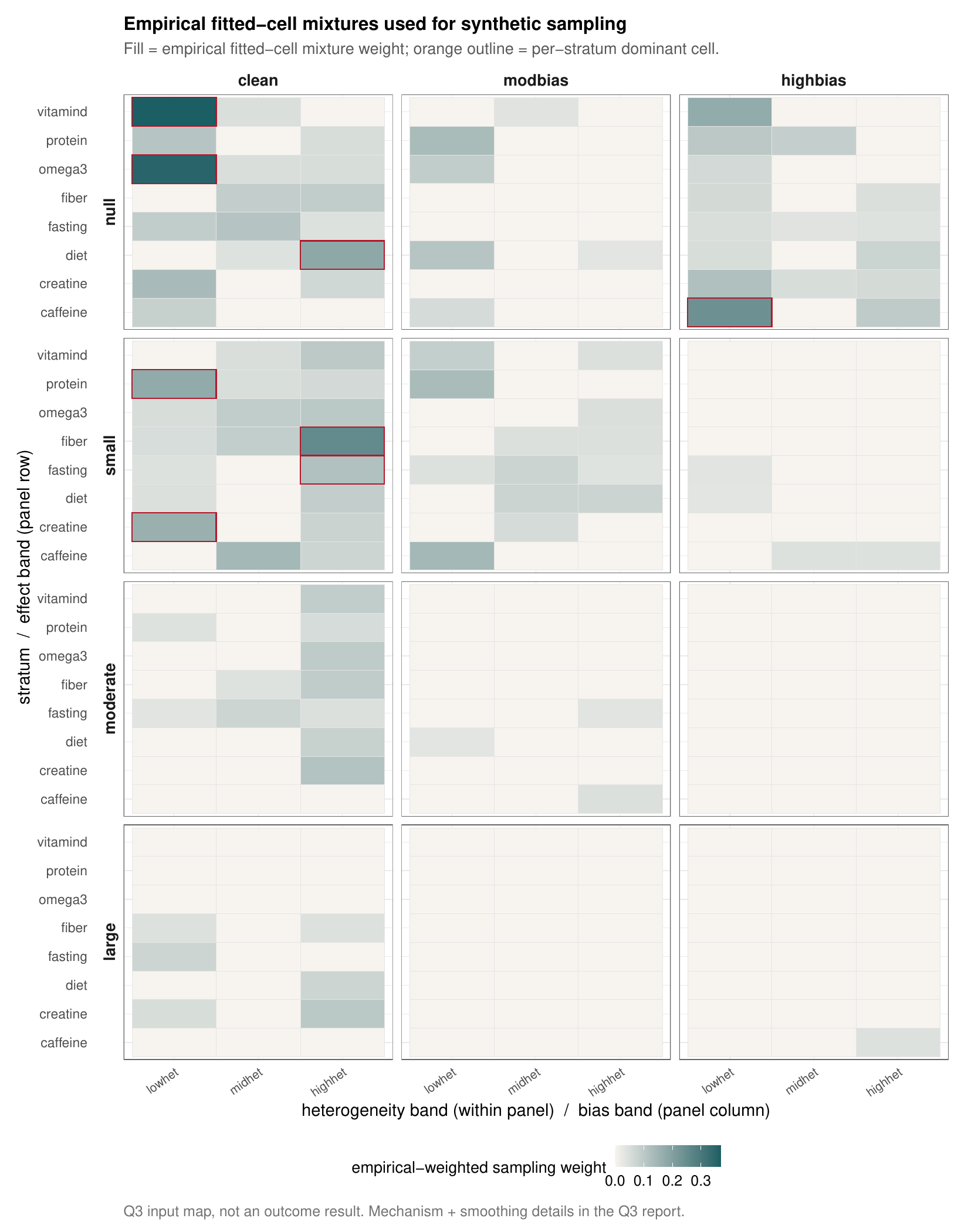}
	\caption{Q3 empirical fitted-profile mixture weights. Each empirical stratum is projected onto the fitted-profile grid using \(|\mu_{\mathrm{BC}}|\), \(\tau_{\mathrm{BC}}\), and \(\log_{10}BF_{\mathrm{bias}}\), then smoothed using support-masked shrinkage weighting (\texttt{eb\_corpus} in the code label). These weights define the primary empirical fitted-profile-weighted synthetic sampling exercise.}
	\label{fig:q3-cell-weights}
\end{figure}

Q3 extends the characterization beyond the exact observed registry by sampling the known-cell synthetic library under empirical fitted-profile mixtures. The first Q3 display is therefore an input map rather than an outcome result. Figure~\ref{fig:q3-cell-weights} shows how the empirical strata are distributed over fitted-profile cells after applying the support-masked shrinkage weighting rule.

This map is useful, but it requires careful interpretation. A large weight on a clean fitted-profile cell does not prove that the underlying empirical literature is free of bias. It means that, under the fitted RoBMA-PSMA ensemble and the available study-level information, the outcome did not produce strong evidence favoring an explicit bias component. In small or weakly informative meta-analyses, absence of bias-component evidence may reflect limited information as much as genuinely clean accumulation. Thus, the fitted-profile map is best read as a description of the empirical audit outputs used to weight the Q3 synthetic sampling exercise, not as a definitive diagnosis of the empirical data-generating process.

\begin{figure}[H]
	\centering
	\includegraphics[width=.9\linewidth]{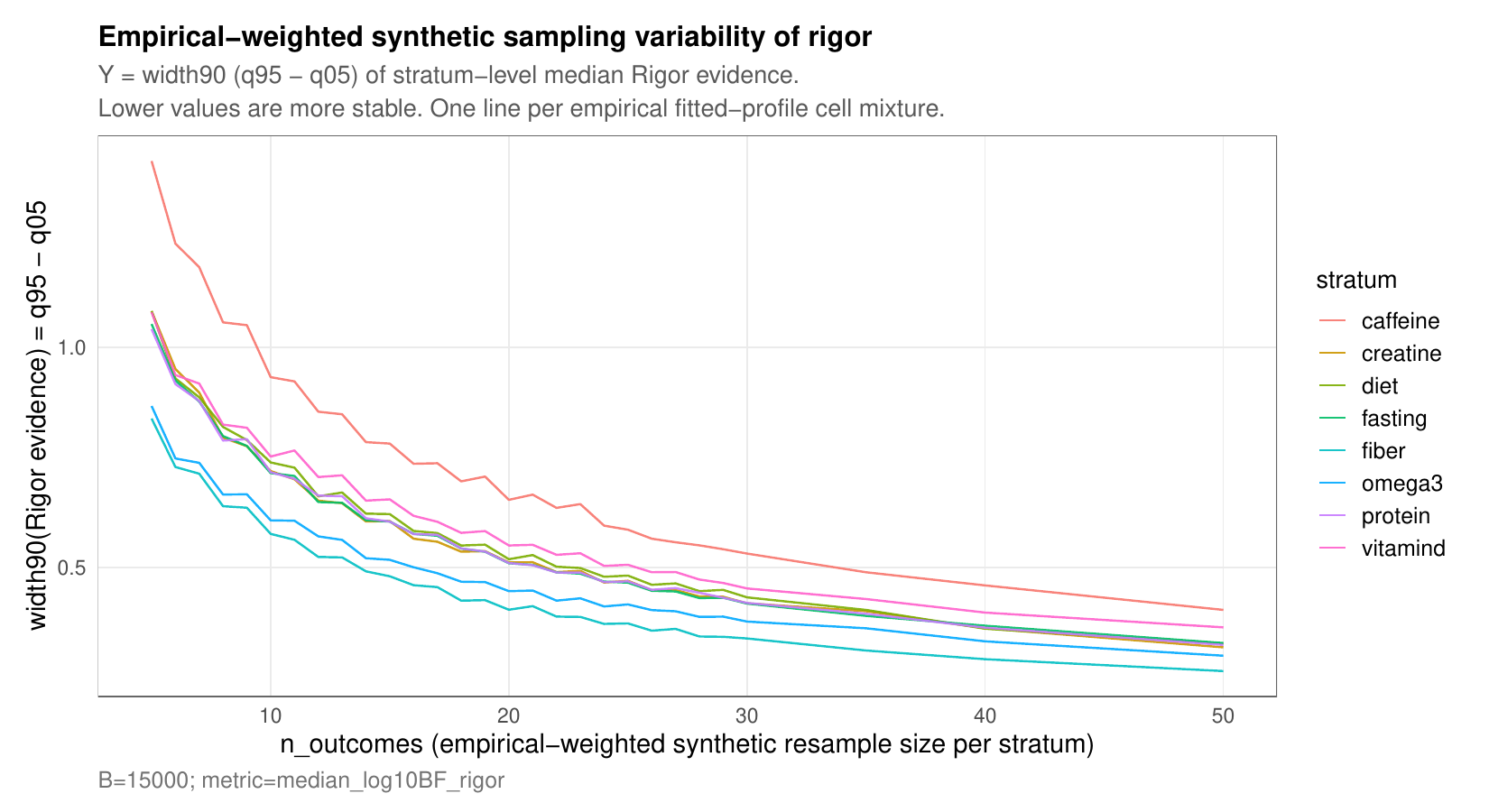}
	\caption{Q3 empirical fitted-profile-weighted synthetic variability for selected rigor. The figure reports \(\mathrm{width90}\) for stratum-level selected rigor when fitted synthetic outcomes are sampled from the Q1 library using empirical fitted-profile weights.}
	\label{fig:q3-rigor-variability}
\end{figure}

Figure~\ref{fig:q3-rigor-variability} then asks how selected-rigor summaries behave when synthetic outcomes are sampled according to those empirical fitted-profile weights. This is the synthetic counterpart to the Q2 empirical resampling stability display. It keeps the empirical fitted-profile mixture fixed and varies \(n_{\mathrm{outcomes}}\), thereby asking how stable the corresponding synthetic profile becomes as more outcome-level meta-analyses are accumulated.

The value of this display is not that it reproduces the empirical registry exactly. Rather, it asks whether the rigor workflow stabilizes under empirical-like profile mixtures when the sampled outcomes come from a broader known-cell synthetic library. This is a portability check: it evaluates whether selected rigor remains interpretable when the empirical profile structure is imposed on synthetic evidence objects beyond the exact finite nutrition registry.

\begin{figure}[H]
	\centering
	\includegraphics[width=\linewidth]{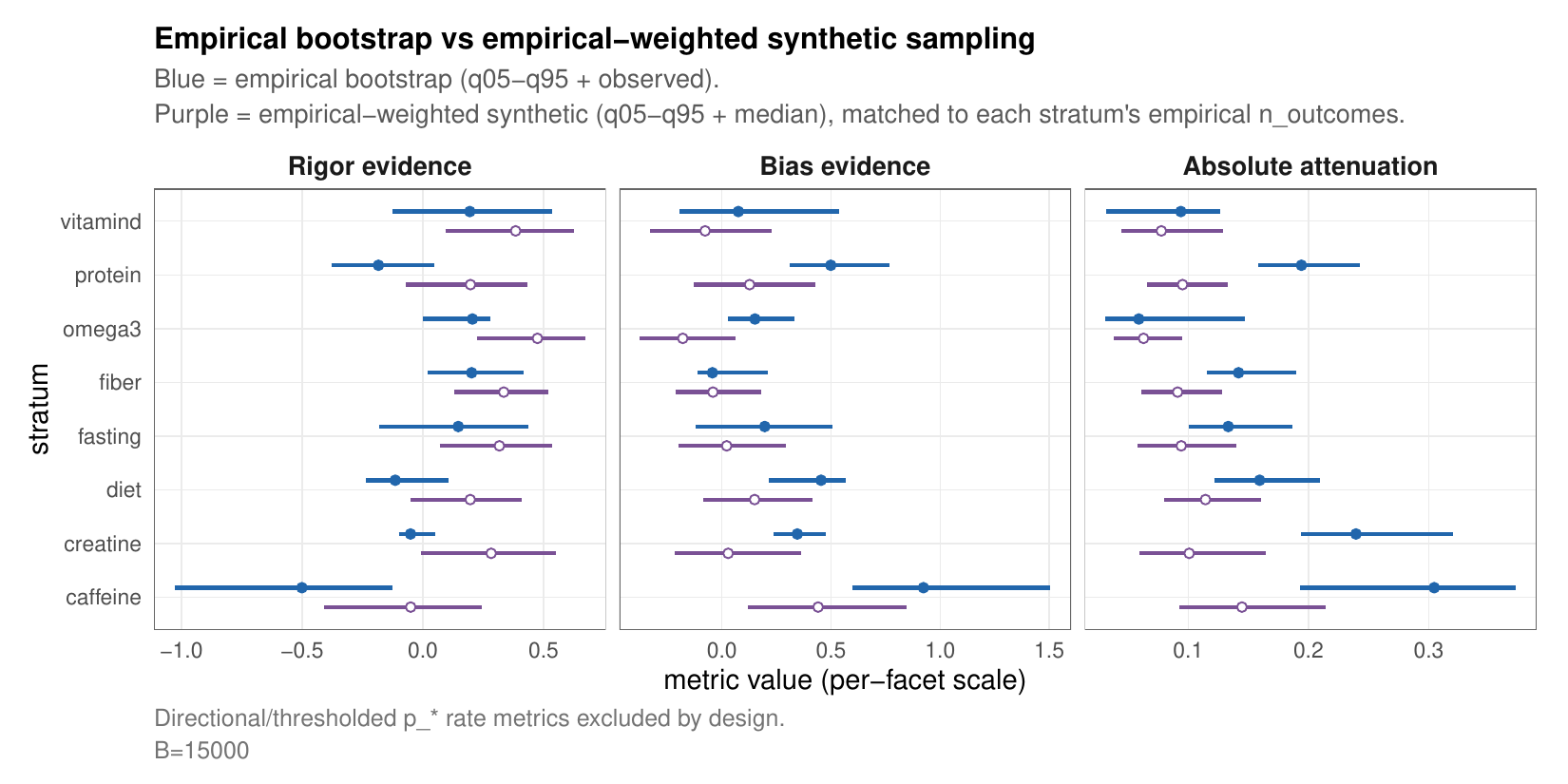}
	\caption{Q3 bridge comparison between empirical bootstrap summaries and empirical fitted-profile-weighted synthetic summaries. The comparison is made at matched empirical stratum sizes for rigor evidence, bias evidence, and absolute attenuation. It is interpreted as a contrast, not as calibration of the synthetic library to the empirical corpus.}
	\label{fig:q3-empirical-synthetic-bridge}
\end{figure}

Finally, Figure~\ref{fig:q3-empirical-synthetic-bridge} compares empirical bootstrap summaries with empirical fitted-profile-weighted synthetic summaries for the three manuscript-facing core metrics. This figure should not be read as a calibration target. The purpose is not to tune the synthetic library until it matches the empirical registry, nor to claim that the empirical registry was generated by the synthetic regimes. Differences between the empirical bootstrap intervals and the empirical-weighted synthetic intervals are informative: they show where the empirical registry has structure that is not fully reproduced by the fitted-profile-weighted synthetic sampling exercise.

As in Q2, the primary interpretive panel is selected rigor. Bias evidence and absolute attenuation are retained as diagnostic companions: bias evidence reflects support for explicit bias components in the fitted ensemble, and attenuation describes baseline-versus-bias-aware movement. These quantities can help diagnose why a stratum behaves as it does, but they are not the primary basis for claims about evidential yield or portability.

Q3 therefore has a narrower interpretation than either calibration or external validation. It does not estimate the true data-generating mixture of the nutrition literature. It asks whether the empirical fitted-profile structure, when imposed on a known-cell synthetic library processed through the same audit workflow, yields stability and core-metric behavior that helps contextualize the empirical resampling results and extends the characterization beyond the exact observed registry.

\subsection{Synthesis of the simulation/resampling characterization}
\label{sec:simulation-synthesis}

The three exercises complete the ADEMP characterization in complementary ways. Q1 evaluates known-cell operating behavior: selected rigor, effect recovery, and attenuation under controlled synthetic effect, heterogeneity, and bias-burden regimes. It provides reassurance that the workflow behaves coherently under the synthetic cells used for this paper, while also showing that weak effects remain difficult to resolve. Q2 evaluates empirical finite-registry stability: whether the observed nutrition summaries are stable under outcome-row and source-cluster resampling. Q3 evaluates empirical-profile-weighted synthetic behavior: whether selected rigor remains stable and interpretable when the known-cell synthetic library is sampled under empirical fitted-profile mixtures.

This characterization supports the use of rigor as a corpus-level evidential-yield estimand in the case study that follows. The simulation layer does not claim that the nutrition corpus is a random sample from all nutrition evidence, and Q3 does not assign empirical outcomes to true DGM cells. Instead, Section~\ref{sec:simulation} specifies aims, mechanisms, estimands, methods, and performance measures for evaluating whether the fixed audit workflow produces interpretable known-cell behavior, stable empirical summaries, and useful empirical-profile-weighted synthetic comparisons.

The practical implication is that selected rigor should be interpreted together with its stability context. Large differences in stratum-level rigor are more persuasive when the relevant width curves and viability summaries indicate stable estimation at the observed outcome counts. Small differences, or differences in regions with broad \(\mathrm{width90}\), should be treated as provisional. Bias evidence and attenuation help explain why rigor is high or low, but they remain companions to the selected-rigor estimand. These are the operating questions needed before using the workflow descriptively on the nutrition intervention meta-analyses in Section~\ref{sec:case-study}.

\section{Case study: nutrition intervention meta-analyses}
\label{sec:case-study}

We now illustrate the workflow with the nutrition demonstration corpus described in Section~\ref{sec:corpus}. The purpose of this section is not to adjudicate nutrition interventions one by one or to estimate population-level properties of nutrition evidence as a field. The case study is a worked demonstration of the evidential-audit workflow applied to a finite, purposive corpus of reconstructed meta-analytic evidence objects. The findings below should therefore be read as descriptive properties of the analyzed audit corpus rather than as sampling-based estimates for the broader nutrition literature.

The case study uses \CorpusOutcomeN{} reconstructed outcome-level meta-analyses across \CorpusStratumN{} intervention strata. Each outcome was represented by audit-ready study-level effects and standard errors, direction-harmonized, fit with the matched Bayesian random-effects baseline, and then fit with the RoBMA-PSMA bias-aware ensemble. The empirical reporting hierarchy follows Sections~\ref{sec:rigor-estimand} and~\ref{sec:simulation}: selected rigor is the headline evidential-yield quantity; effect and modeled-bias evidence explain the fitted evidence profile; attenuation from the matched baseline is a companion contrast rather than the main result.

\subsection{Case-study corpus and reporting target}

The empirical target in this section is the evidence profile produced by applying the common audit engine to the reconstructed nutrition corpus. The relevant unit is an outcome-level meta-analytic evidence object after reconstruction and common reanalysis, not a source article, an intervention category, or a source-paper pooled estimate. This distinction is central to the case study: the workflow is designed to ask what a corpus of fitted evidence objects yields under a shared evidential architecture.

The primary case-study question is how much clean resolved evidence the analyzed corpus produces. Here, ``clean'' has the model-space meaning used throughout the paper: the selected branch of the fitted RoBMA-PSMA ensemble does not contain an explicit publication-selection or small-study-effect component. Clean evidence therefore does not mean that a literature is free of all bias; it means that the fitted ensemble updated support toward a resolved effect or no-effect branch that did not require the specified explicit bias-adjustment component. Component evidence and matched-baseline attenuation are then used to explain the resulting rigor profile.

\subsection{Headline selected-rigor distribution across strata}

\begin{figure}[tbp]
	\centering
	\includegraphics[width=\linewidth,height=0.78\textheight,keepaspectratio]{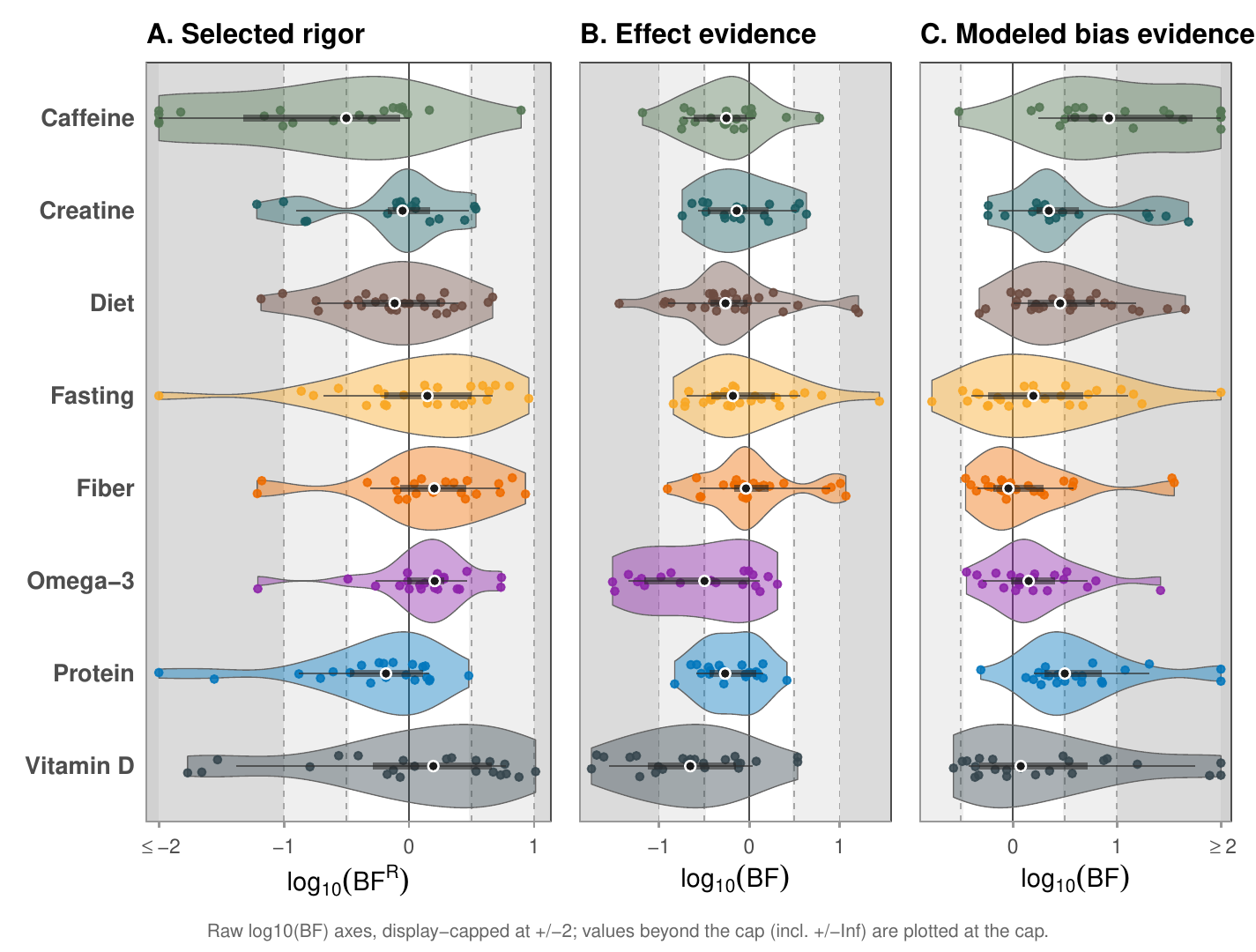}
	\caption{Outcome-level evidence distributions across nutrition strata for the main case-study quantities. Panel A shows selected rigor, the joint model-family Bayes-factor estimand for clean resolved evidence (\(\log_{10}BF_k^R\)); Panel B shows marginal effect-component evidence and Panel C marginal modeled-bias evidence (each a component inclusion \(\log_{10}\) BF). Within each stratum, violins show the outcome-level distribution; points show individual reconstructed outcomes, the central marker shows the median, the thick bar the interquartile range, and whiskers the 10--90\% range. Displayed values are capped at \(|\log_{10}(\mathrm{BF})|=2\) for plotting only; raw registry/sidecar values are unchanged.}
	\label{fig:violin_stack_fullpage}
\end{figure}

Figure~\ref{fig:violin_stack_fullpage} is the main evidence display for the case study. Panel A shows the headline quantity: selected rigor, \(\log_{10}BF_k^R\), across reconstructed outcomes and nutrition strata. Panels B and C show the two main evidence companions displayed in the main text: marginal effect-component evidence and marginal modeled-bias evidence. This arrangement keeps the visual emphasis on selected rigor while making the principal explanatory evidence dimensions visible in the same display.

Panel A should be read as a corpus-audit distribution, not as a ranking of nutrition interventions. Each point is a reconstructed outcome-level evidence object, and each violin summarizes the distribution of those outcomes within a stratum. Positive selected-rigor values indicate movement toward the better-supported clean resolved branch. Values near zero indicate little net support for clean resolved evidence. Negative values indicate that even the better-supported clean branch lost support relative to its complement. Negative rigor is therefore a failure of clean evidential resolution, not evidence for no effect; the selected branch label is interpretable only together with the sign and magnitude of \(\log_{10}BF_k^R\).

The case-study pattern is mixed rather than uniformly reassuring. Some outcomes and strata show positive clean evidential support, but many outcomes cluster near zero or below zero. The workflow therefore does not simply reproduce conventional pooled-effect magnitudes or count nominally nonzero results. It asks whether the reconstructed evidence supports a resolved effect or no-effect conclusion in the part of the fitted model space that does not require the explicit modeled-bias component.

\subsection{Component evidence and corpus-level evidential patterns}

Table~\ref{tab:nutrition_overview_diagnostics} gives the corresponding numerical summary for the audit corpus. The clearest corpus-level pattern is limited effect-component resolution. Most reconstructed outcomes fell in the weak or inconclusive region for effect evidence, while strong evidence favoring a nonzero effect was comparatively uncommon. This pattern is central to the case study: many reconstructed outcomes that enter the workflow as published meta-analytic evidence objects do not yield strong model-family evidence for a nonzero effect after bias-aware model averaging.

Modeled-bias evidence provides an important explanation for the selected-rigor profile. A substantial subset of outcomes showed at least moderate evidence for explicit bias-adjustment components, and many of these same outcomes remained weak or inconclusive for the effect component. This joint pattern is precisely the kind of result the workflow is designed to surface: the observed study record can favor explicit bias-adjustment structure without producing strong evidence for a nonzero effect.

Heterogeneity remains an important part of the fitted evidence profile even though it is not a main-text visual panel in Figure~\ref{fig:violin_stack_fullpage}. Table~\ref{tab:nutrition_overview_diagnostics} reports the heterogeneity rows alongside the effect, modeled-bias, joint-pattern, and conventional-magnitude summaries. Reporting these dimensions separately is a useful feature of the audit workflow: weak effect evidence, modeled-bias evidence, and heterogeneity evidence are distinct fitted-model patterns, even when all three may contribute to low clean evidential resolution.

\begin{table}[!htbp]
	\centering
	\small
	\caption{Audit-corpus evidential patterns surfaced by the workflow. All percentages are descriptive properties of the demonstration corpus rather than estimates of population parameters of the broader nutrition literature. Bayes factors are reported on a \(\log_{10}\) scale; thresholds at \(0.5\) and \(1\) correspond to \(\mathrm{BF}\approx 3.16\) and \(10\), respectively.}
	\label{tab:nutrition_overview_diagnostics}
	\begin{tabularx}{\linewidth}{@{}l l r r X@{}}
		\toprule
		\textbf{Component} & \textbf{Criterion} & $\boldsymbol{n/N}$ & \textbf{\%} & \textbf{Interpretation} \\
		\midrule
		
		\multicolumn{5}{@{}l}{\textit{Effect evidence}}\\[-2pt]
		& $|\log_{10}\mathrm{BF}_{\text{effect}}|\le 0.5$ & \EffectInconclusiveN/\CorpusOutcomeN & \EffectInconclusivePct\% & Weak / inconclusive evidence for an effect component. \\
		& $\log_{10}\mathrm{BF}_{\text{effect}}>0.5$ & \EffectModeratePosN/\CorpusOutcomeN & \EffectModeratePosPct\% & At least moderate evidence favoring a nonzero effect. \\
		& $\log_{10}\mathrm{BF}_{\text{effect}}>1$ & \EffectStrongPosN/\CorpusOutcomeN & \EffectStrongPosPct\% & Strong evidence favoring a nonzero effect. \\
		& $\log_{10}\mathrm{BF}_{\text{effect}}<-0.5$ & \EffectModerateNegN/\CorpusOutcomeN & \EffectModerateNegPct\% & At least moderate evidence favoring a null-effect component. \\
		& $\log_{10}\mathrm{BF}_{\text{effect}}<-1$ & \EffectStrongNegN/\CorpusOutcomeN & \EffectStrongNegPct\% & Strong evidence favoring a null-effect component. \\
		\addlinespace[2pt]
		
		\multicolumn{5}{@{}l}{\textit{Bias-adjustment evidence}}\\[-2pt]
		& $\log_{10}\mathrm{BF}_{\text{bias}}>0.5$ & \BiasModeratePosN/\CorpusOutcomeN & \BiasModeratePosPct\% & At least moderate evidence for explicit bias-adjustment components. \\
		& $\log_{10}\mathrm{BF}_{\text{bias}}>1$ & \BiasStrongPosN/\CorpusOutcomeN & \BiasStrongPosPct\% & Strong evidence for explicit bias-adjustment components. \\
		\addlinespace[2pt]
		
		\multicolumn{5}{@{}l}{\textit{Heterogeneity evidence}}\\[-2pt]
		& $\log_{10}\mathrm{BF}_{\text{het}}>0.5$ & \HetModeratePosN/\CorpusOutcomeN & \HetModeratePosPct\% & At least moderate evidence for between-study heterogeneity. \\
		& $\log_{10}\mathrm{BF}_{\text{het}}>1$ & \HetStrongPosN/\CorpusOutcomeN & \HetStrongPosPct\% & Strong evidence for heterogeneity. \\
		\addlinespace[2pt]
		
		\multicolumn{5}{@{}l}{\textit{Joint pattern}}\\[-2pt]
		& \makecell[l]{$\log_{10}\mathrm{BF}_{\text{bias}}>0.5$\\
			and $|\log_{10}\mathrm{BF}_{\text{effect}}|\le 0.5$}
		& \JointBiasWeakEffectN/\CorpusOutcomeN & \JointBiasWeakEffectPct\% & Modeled-bias evidence despite weak effect evidence. \\
		\addlinespace[2pt]
		
		\multicolumn{5}{@{}l}{\textit{Conventional magnitude subset}}\\[-2pt]
		& $|\mu_{\mathrm{RE}}|\ge 0.2$ & \ClaimedEffectN/\CorpusOutcomeN & \ClaimedEffectPct\% & Outcomes with conventionally nontrivial baseline random-effects estimates. \\
		& Shrinkage $\ge 50\%$ & \ClaimedShrinkFiftyN/\ClaimedEffectN & \ClaimedShrinkFiftyPct\% & Large attenuation under RoBMA-PSMA. \\
		& \makecell[l]{$|\log_{10}\mathrm{BF}_{\text{effect}}|\le 0.5$\\
			(within $|\mu_{\mathrm{RE}}|\ge 0.2$)}
		& \ClaimedEffectInconclusiveN/\ClaimedEffectN & \ClaimedEffectInconclusivePct\% & Most conventionally meaningful baseline effects remained evidentially inconclusive. \\
		\bottomrule
	\end{tabularx}
\end{table}

\subsection{Matched-baseline attenuation as a companion result}

The matched Bayesian random-effects baseline provides a standardized comparator for interpreting how bias-aware model averaging changes estimated effects. Because the baseline random-effects fit and the RoBMA-PSMA fit use the same audit-ready data object for each outcome, the contrast is interpretable as baseline-versus-bias-aware movement under the workflow rather than as an artifact of heterogeneous source-paper estimators, software, or modeling conventions.

The conventional-magnitude rows of Table~\ref{tab:nutrition_overview_diagnostics} summarize this companion pattern. Among outcomes with matched-baseline estimates of at least moderate standardized magnitude, large attenuation under RoBMA-PSMA was common, and many of those same outcomes remained weak or inconclusive for effect evidence. The important point is not simply that bias-aware estimates were often smaller. Rather, many outcomes that would attract attention under matched-baseline summaries did not translate into strong clean evidential support after model uncertainty and explicit bias-adjustment components were propagated through the workflow.

This attenuation result should remain in its companion role. It helps explain why nominally meaningful baseline estimates may lose clean evidential support, but it is not the evidential-yield estimand. A corpus can accumulate useful evidence by supporting effects or by supporting no-effect conclusions. Attenuation alone cannot distinguish those cases; selected rigor is designed to preserve that distinction. Per-stratum paired baseline and bias-aware estimate displays are provided in the Supplementary Material.

\subsection{Case-study synthesis and stability context}

The nutrition case study illustrates the intended use of the workflow as a corpus-scale evidential audit. In this finite, purposive corpus, selected rigor shows limited clean evidential resolution across many outcome-level meta-analyses. The component evidence explains that profile: strong evidence for nonzero effects was uncommon, modeled-bias evidence was frequent enough to matter, and heterogeneity evidence was substantial in many outcomes. The conventional-magnitude subset further shows that many baseline effects of at least moderate standardized size were strongly attenuated or remained evidentially weak after the bias-aware ensemble was fit.

These findings should be interpreted with the finite-registry stability context from Section~\ref{sec:q2-empirical-resampling}. The empirical resampling layer asks how sensitive stratum summaries are to outcome-row and source-cluster resampling, and Section~\ref{sec:q3-empirical-weighted-synthetic} extends that characterization with empirical fitted-profile-weighted synthetic sampling. Those exercises do not make the nutrition corpus representative of all nutrition evidence, and they do not turn fitted-profile cells into true data-generating labels. They provide stability context for reading the descriptive case-study summaries.

Additional stratum-level plots, outcome-level diagnostics, paired baseline-versus-bias-aware estimate displays, and the full outcome registry are provided in the Supplementary Material. Those materials support inspection of individual evidence objects and strata, while the main text keeps the emphasis on the reusable workflow: selected rigor as the headline evidential-yield estimand, component model-family evidence as the explanatory decomposition, and attenuation as a matched-baseline companion rather than the central result.

\section{Discussion}
\label{sec:discussion}

\subsection{Main contribution}

This paper provides a reusable framework for measuring evidential yield in meta-analytic corpora. The workflow converts reconstructed or supplied study-level meta-analytic evidence objects into a common Bayesian model-family evidence registry: inputs are standardized, fitted with a matched Bayesian random-effects baseline and a bias-aware Bayesian model-averaged ensemble, reduced to auditable sidecars, and summarized through component and joint model-family evidence. This structure makes it possible to compare evidence accumulation across strata, journals, research programs, funder portfolios, or prospective evidence streams using a declared grouping scheme and a common evidential contract.

The central methodological object is the rigor estimand. Rigor is a joint model-family Bayes factor for the better-supported clean resolved branch of the fitted ensemble, where clean denotes absence of an explicit publication-selection or small-study-effect component in the model space. This definition changes the target of corpus comparison. Instead of counting positive estimates, statistically significant findings, or large unadjusted pooled effects, the workflow asks whether a corpus produces resolved evidence for either an effect or no effect in the part of the fitted model space that does not require the specified explicit modeled-bias component. Evidence for no effect can therefore count as successful evidence accumulation, whereas inconclusive, unstable, or bias-dependent evidence receives little rigor support.

This emphasis on joint model-family evidence distinguishes rigor from the companion quantities reported by the workflow. Marginal effect evidence, heterogeneity evidence, modeled-bias evidence, and baseline-to-bias-aware attenuation remain important because they explain the fitted evidence profile. They are not substitutes for rigor. A large marginal effect Bayes factor can coexist with support for explicit bias-adjustment components; a large attenuation contrast can indicate movement in an estimated effect without resolving whether the corpus has accumulated clean evidence; and a selected branch label is only meaningful alongside the sign and magnitude of selected rigor. The reporting hierarchy is therefore deliberate: selected rigor is the headline evidential-yield estimand, component evidence explains why rigor is high or low, attenuation provides a matched-baseline contrast, and simulation/resampling summaries characterize stability.

The workflow builds on RoBMA-PSMA as an existing bias-aware model-averaged audit engine \citep{Maier2023,Bartos2023}. Its novelty is in the corpus-scale protocol wrapped around that engine: the matched-baseline design, the joint rigor estimand, the sidecar/registry reporting contract, the ADEMP-framed simulation/resampling characterization, and the reproducible companion repository. The paper does not ask readers to accept a bespoke synthesis model for one nutrition question. It offers a transparent way to audit evidential yield across many meta-analytic evidence objects using a common model space and a declared grouping scheme.

\subsection{Relationship to Bayesian workflow and bias adjustment}

The workflow is complementary to Bayesian workflow approaches \citep{Gelman2020BayesianWorkflow} for developing and checking individual bias-adjusted meta-analytic models. For example, \citet{JungAloe2026} emphasize prior predictive checks, posterior predictive checks, prior sensitivity, model comparison, and targeted simulation for a study-level risk-of-bias adjustment model. That style of workflow is essential when the goal is to build, justify, and diagnose a particular Bayesian model for a particular synthesis problem.

The present paper addresses a different level of analysis. It holds a documented bias-aware model-averaged ensemble fixed as an audit engine and asks how its model-family evidence can be converted into reusable corpus-level summaries. In this setting, the central challenge is not only whether one adjusted model fits one meta-analysis well. It is whether many outcome-level evidence objects can be processed under a common evidential contract so that effect evidence, no-effect evidence, heterogeneity evidence, modeled-bias evidence, and clean resolved evidence can be compared across strata, source articles, research programs, journals, funder portfolios, or prospective evidence streams.

Model-family Bayes factors are useful because they make the relevant inferential objects explicit. The workflow does not need to choose between an adjusted and unadjusted analysis using only a predictive-information criterion or a qualitative diagnostic. Instead, it reports evidence for the effect component, evidence for heterogeneity, evidence for explicit modeled-bias components, and the joint evidence needed for rigor. Those quantities do not eliminate the need for model checking, prior sensitivity analysis, or substantive judgment. They provide a coherent evidential layer that can be reported alongside those checks and reused across a corpus.

\subsection{What the simulation and case study add}

The simulation/resampling layer is a necessary part of the method rather than a decorative appendix. A rigor estimand is only useful if its behavior is understandable under plausible operating conditions and if corpus summaries are stable enough to interpret. Section~\ref{sec:simulation} therefore characterizes the workflow through three linked ADEMP-framed exercises. Q1 studies known-cell synthetic behavior under controlled effect, heterogeneity, and bias-burden conditions. Q2 studies finite-registry stability in the observed nutrition corpus under outcome-row and source-cluster resampling. Q3 studies empirical fitted-profile-weighted synthetic sampling, connecting empirical fitted-profile structure to the known-cell synthetic library without treating fitted-profile cells as latent data-generating labels.

These exercises have different inferential roles. Q1 provides a local operating check for the assembled workflow under the chosen synthetic design; it is not a full revalidation of RoBMA-PSMA from first principles. Q2 evaluates how sensitive the observed case-study summaries are to finite-registry discreteness, outcome multiplicity, and source-article clustering; it does not convert the purposive nutrition corpus into a population sample. Q3 extends the characterization beyond the exact empirical registry by sampling synthetic outcomes under empirical fitted-profile weights; it is not calibration and does not imply that empirical outcomes were generated from the synthetic cells. Together, the three exercises give readers a stability and behavior context for interpreting selected rigor, modeled-bias evidence, and attenuation.

The nutrition case study then shows what the workflow surfaces in a real, heterogeneous, and substantively familiar corpus. The main result is not a new nutrition consensus statement. It is a demonstration of corpus-scale evidential auditing: selected rigor shows limited clean evidential resolution across many outcome-level meta-analyses; component evidence shows that strong nonzero-effect evidence is uncommon, modeled-bias evidence is frequent enough to matter, and heterogeneity is substantial in many outcomes; and the conventional-magnitude subset shows that many baseline effects of at least moderate standardized size are strongly attenuated or remain evidentially weak after the bias-aware ensemble is fit. These are descriptive properties of the analyzed audit corpus, not population-level estimates for nutrition evidence as a field.

This case-study role is important. A methods paper can define an estimand and describe a pipeline, but readers also need to see how the estimand changes interpretation in an empirical corpus. In the nutrition demonstration, attenuation alone would be an incomplete story: it would describe how estimated effects move under bias-aware modeling, but not whether the evidence resolves toward an effect or no effect without explicit modeled-bias support. Rigor supplies that missing evidential target, while the component Bayes factors and attenuation rows explain the mechanisms behind the rigor profile.

\subsection{Reproducibility and reusable infrastructure}

The public companion repository is part of the methodological contribution. Corpus-scale evidential auditing is vulnerable to hidden degrees of freedom: reconstruction choices, effect-direction conventions, model-fitting configuration, inclusion/exclusion rules, aggregation weights, figure scripts, and manuscript tables can drift apart unless the workflow enforces a reporting contract. The repository is designed to reduce that risk. It separates expensive model fitting from downstream summarization, stores accepted outcome-level sidecars, rebuilds a canonical registry, and generates tables and figures from registry-derived quantities rather than from hand-entered results.

This design supports reproducibility in the ordinary sense: readers can inspect the analysis-ready inputs, reconstruction records, scripts, sidecars, registry files, generated tables, figures, and documentation used to produce the manuscript. It also supports reuse. A future analyst can adapt the same input contract, sidecar schema, registry rebuild, aggregation rules, and reporting layer to a new corpus without redesigning the full workflow. The nutrition case study and simulation subproject are therefore examples of the workflow, not the boundary of the workflow.

For methods readers, this infrastructure matters because the estimand alone is not enough. A joint rigor Bayes factor can be defined mathematically, but corpus-scale use requires a durable implementation: stable file paths, generated artifacts, versioned numbers, reproducible visual outputs, and clear documentation of what is fitted once versus what is recomputed downstream. The repository also makes explicit which parts of the process are general and which are corpus-specific. Reconstruction and extraction decisions are necessarily corpus-specific; the audit engine, sidecar reduction, registry contract, rigor calculation, and reporting layer are intended to be reusable.

\subsection{Portability and intended uses}

The workflow is portable to any analyst-defined corpus whose outcome-level meta-analytic evidence objects can be represented by study-level estimates and standard errors on a common analysis scale. Potential schemes include intervention classes, journal-year windows, field-year windows, source articles, funder portfolios, research programs, clinical or policy domains, or continuously updated evidence streams. The grouping scheme defines the comparison; the fitted outcome-level evidence records remain the basic units.

This portability should be understood as a bounded methodological claim. The workflow does not replace primary systematic review, does not certify study-level internal validity, does not prove that any adjusted estimate is unbiased, and does not remove the need for domain expertise. It supplies a standardized evidential layer for asking whether a declared corpus is producing clean resolved evidence. That layer can sit alongside conventional systematic-review judgments, risk-of-bias assessments, certainty-of-evidence frameworks, and substantive interpretation.

One natural use is retrospective audit. A mature literature may contain many published syntheses, but those syntheses may not reveal whether the evidence is actually resolving. Applying the workflow to a corpus of legacy meta-analyses can identify strata where clean evidence has accumulated, strata where evidence remains weak or bias-dependent, and strata where source-article multiplicity or outcome clustering makes summaries unstable. This is useful for deciding where replication, larger trials, better reporting, or better-designed syntheses are most needed.

A second use is prospective evidence monitoring. As new meta-analyses appear, author-supplied or reconstructed study-level inputs could be passed through the same audit engine and added to a registry. Stratum- or program-level rigor summaries could then be updated over time. In that setting, rigor becomes a measure of evidential yield: not whether a program produces many positive findings, but whether its cumulative synthesis activity produces durable evidence for effects or for their absence without requiring explicit modeled-bias components. This is especially relevant for journal, funder, and program-level evaluation, where a positive-finding count would reward exactly the wrong target.

\subsection{Limitations}

Several limitations should guide interpretation. First, the nutrition corpus is purposive and extractability-conditioned. Its summaries are descriptive properties of the analyzed audit corpus and should not be generalized as population estimates for nutrition as a field. The same limitation will apply to any corpus assembled without a sampling design that justifies population inference.

Second, reconstruction introduces uncertainty. The workflow can accept already-standardized study-level inputs, but legacy meta-analyses often require conversions, standard-error reconstruction, direction harmonization, and assumptions about repeated-measures or crossover designs. These steps are documented as part of the audit trail, but they are not lossless. The workflow makes such decisions visible; it does not make them disappear.

Third, the common effect-size and direction conventions are necessary for corpus comparison but require substantive judgment. A positive direction can be defined differently across outcomes, and some outcomes may not map neatly onto a single beneficial direction. Direction harmonization is a coordinate-system choice, not a claim that all outcomes share the same substantive meaning.

Fourth, the workflow is model-space dependent. Clean evidence means support for a branch of the fitted RoBMA-PSMA ensemble without the specified explicit publication-selection or small-study-effect component. It does not mean that the underlying literature is free of all bias. RoBMA-PSMA does not directly encode study-level risk-of-bias domains such as allocation concealment, blinding, attrition, or selective outcome reporting unless those features are introduced through additional model components or metadata. The bias component should therefore be interpreted as evidence for the modeled publication-selection and small-study-effect structures, not as a complete bias diagnosis.

Fifth, rigor is an evidential quantity, not a truth label. A large positive selected rigor value indicates that the observed data updated support toward the better-supported clean resolved branch under the specified ensemble. It does not prove that an effect is true, that no effect is true, or that all bias mechanisms are absent. Conversely, negative rigor means clean resolved evidence is disfavored; it is not evidence for no effect. The selected branch label should therefore be reported as interpretive metadata and read only alongside the sign and magnitude of selected rigor.

Sixth, the matched baseline is a standardized comparator rather than a privileged truth model. It is included so that baseline-to-bias-aware attenuation can be interpreted across the corpus without confounding the comparison with heterogeneous source-paper estimators. Other baseline choices could be substituted in future implementations. Similarly, attenuation is not a success metric: it explains movement from the matched baseline to the bias-aware ensemble, but it does not measure clean evidential yield.

Seventh, the simulation/resampling characterization is conditional on the mechanisms studied. Q1 depends on the chosen synthetic design; Q2 studies stability of the observed finite registry rather than a known population; and Q3 uses empirical fitted-profile-weighted synthetic sampling rather than a new empirical data-generating mechanism. The performance summaries should therefore be read as behavior and stability characterizations for the workflow under those mechanisms, not as universal operating guarantees.

Eighth, the current implementation focuses on RoBMA-PSMA as the audit engine. This is an advantage for comparability because all outcomes share a common model space, but it also means that some findings may be specific to the RoBMA-PSMA ensemble, its priors, and its component definitions. Comparator analyses with alternative bias-adjustment families and systematic prior sensitivity analyses remain important extensions.

Ninth, outcome multiplicity and source clustering are only partially addressed by article-balanced summaries and source-cluster resampling. These tools make multiplicity visible and reduce the influence of prolific source syntheses in summaries, but they are not a full hierarchical model of dependence among outcomes, source articles, interventions, and primary studies. Future corpus-audit designs may need richer dependence models when the inferential target requires them.

These limitations bound the claims of the paper but do not weaken its central methodological contribution. The workflow provides a transparent protocol for turning many meta-analytic evidence objects into comparable model-family evidence summaries, with rigor as the headline estimand and simulation/resampling as a stability layer. The case study shows what that protocol reveals in one demanding corpus; the repository makes the protocol inspectable, reproducible, and adaptable.

\subsection{Future directions}
\label{sec:rigor-implementation}

Several extensions follow naturally. First, the simulation/resampling characterization can be expanded beyond the present ADEMP layer. The current design evaluates known-cell behavior, finite-registry stability, and empirical fitted-profile-weighted synthetic sampling for the assembled workflow. Future work can extend this to additional synthetic mechanisms, alternative study-count distributions, more severe selection regimes, different heterogeneity structures, and richer outcome-dependence patterns. The goal would be to map the operating behavior of selected rigor and its companion summaries across a wider range of corpus geometries.

Second, systematic prior sensitivity should be incorporated into the reporting contract. Because rigor is a joint model-family Bayes factor, it can be sensitive to prior model probabilities and parameter priors in ways that are not fully visible from marginal effect or bias-component summaries alone. A reusable audit workflow should therefore expose prior-sensitivity variants for selected rigor, branch-specific rigor, modeled-bias evidence, and attenuation, while keeping the primary analysis pre-specified.

Third, comparator bias-adjustment analyses would clarify how much of the corpus-level signal is specific to RoBMA-PSMA. PET-PEESE, weight-function selection models, trim-and-fill, Copas-type selection models, or other bias-adjustment approaches could be run as structured comparators where feasible. These analyses should not replace the joint rigor estimand unless the comparator model space can support analogous joint model-family evidence, but they can help distinguish conclusions that are robust across adjustment families from conclusions that depend strongly on the chosen audit engine.

Fourth, the model space could be extended to include additional bias or quality dimensions. Study-level risk-of-bias information, funding indicators, preregistration status, outcome-reporting indicators, journal features, or design-quality metadata could be incorporated into future bias-aware ensembles or into downstream explanatory models. Such extensions would move beyond publication-selection and small-study-effect components toward a broader evidential-quality architecture, but they would need to preserve the core principle that the estimand must be defined from explicit model-family probabilities rather than from ad hoc diagnostic labels.

Fifth, the workflow should be applied to additional corpora. The most informative next applications would not simply add more nutrition topics. They would test portability under different corpus schemes: journal windows, field-year windows, funder portfolios, clinical guideline domains, education or psychology corpora, and prospectively updated evidence streams. Such applications would help determine whether selected rigor behaves as a useful comparative measure of evidential yield outside the motivating case study.

Sixth, the repository can be developed into a more general toolkit for evidential audits. The current companion repository documents the pipeline, inputs, sidecars, registry, generated tables, figures, and simulation layer. Future versions could provide template schemas for new corpora, containerized environments, validation checklists, example minimal datasets, standardized visual/reporting modules, and automated release manifests. These additions would make the workflow easier for other research-synthesis groups to reuse without weakening the current emphasis on explicit contracts and inspectable artifacts.

Finally, there is a downstream synthesis problem. Outputs from the audit---bias-aware estimates, posterior summaries, component Bayes factors, branch-specific rigor Bayes factors, and selected rigor---could become inputs to higher-level Bayesian evidence synthesis, evidence-quality modeling, or research-program evaluation. That integration is not automatic. It raises problems of scale matching, dependence, double use of primary trials, prior construction, and the choice of which audit outputs should be propagated forward. The present paper performs the audit step; developing principled downstream uses of audit outputs is a separate methodological program.

\subsection{Conclusion}

Rigor operationalizes one aspect of productive evidence accumulation: whether a corpus of meta-analytic evidence converts accumulated studies into resolved support for effects or no effects without relying on explicit modeled publication-selection or small-study-effect components. By defining rigor as a joint model-family Bayes factor, embedding it in a reproducible audit workflow, characterizing the workflow through simulation/resampling, and demonstrating it in a purposive nutrition corpus, this paper provides a portable framework for evidential auditing of meta-analytic corpora. The companion repository is essential to that framework because it turns the manuscript from a conceptual proposal into an inspectable and reusable workflow. The broader aim is to help research synthesis move from reporting many pooled estimates toward measuring whether evidence accumulation itself is producing clean, durable evidential resolution.

\section*{Acknowledgments}
The author thanks Dr.~Hassan Elsalloukh and Franti\v{s}ek Barto\v{s} for helpful comments on earlier drafts of the manuscript.

\section*{Generative AI disclosure}
Generative AI tools were used extensively as assistive tools during this project. OpenAI ChatGPT (including GPT-5.5 Thinking sessions) and Anthropic Claude were used through their standard user interfaces during May 2026 to support manuscript drafting and revision, code and workflow review, LaTeX and document-organization tasks, repository-documentation cleanup, release planning, and preparation of reproducibility materials. The tools were also used to help identify inconsistencies across manuscript, supplement, and repository documentation.

The AI tools did not generate the study-level data, perform the RoBMA-PSMA model fitting, execute the simulation/resampling runs, make autonomous scientific decisions, or serve as a source of uncited factual claims. All AI-assisted text, code suggestions, workflow recommendations, and release materials were reviewed, edited, and verified by the author, who takes full responsibility for the analysis, interpretation, and final content.

\section*{Author contributions}
Matt Hester: Conceptualization, Methodology, Software, Formal analysis, Data curation, Visualization, Writing---original draft, Writing---review and editing

\section*{Funding statement}
No specific funding was received for this work.

\section*{Competing interests}
The author declares no competing interests.

\newpage
\section*{Data availability statement}
All analysis-ready empirical datasets, extraction workbooks documenting their derivation from the source meta-analyses, R scripts implementing the pipeline, empirical sidecars, derived reporting tables, figure-generation code, simulation source and design files, and accompanying documentation are openly available in the versioned companion repository, archived as v1.0.0 on Zenodo \citep{HesterEvidentialAuditWorkflow} at \url{https://doi.org/10.5281/zenodo.20467258}. The live development repository is at \url{https://github.com/matthewahester/evidential-audit-workflow}. Generated simulation datasets, raw simulation draws, and full fitted simulation output trees are omitted from the archived release because they are regenerable from the archived scripts, design grid, and recorded computational environment; the materials provided are sufficient to regenerate the reported empirical summaries directly and to reproduce the simulation displays through the documented re-run procedure.

\section*{Ethics statement}
Not applicable. This study involved reanalysis of published meta-analytic datasets and did not involve new data collection from human participants, identifiable personal data, or animal experimentation.

\section*{Permissions}
Not applicable. Figures and tables were generated by the author from the described workflow and archived materials.

\section*{Patient consent and clinical trial registration}
Not applicable.
	
	\clearpage
%
%
\nocite{chilibeckEffectCreatineSupplementation2017,desaiEffectCreatineSupplementation2024,devriesCreatineSupplementationResistance2014,domaParadoxicalEffectCreatine2022,fernandez-landaEffectsCreatineMonohydrate2023,kazeminasabEffectsCreatineSupplementation2025,lanhersCreatineSupplementationLower2015,lanhersCreatineSupplementationUpper2017,ajalaSystematicReviewMetaanalysis2013,brand-millerLowGlycemicIndex2003,buenoVerylowcarbohydrateKetogenicDiet2013,firthEffectsDietaryImprovement2019,greylingAcuteGlycemicInsulinemic2020,huangVegetarianDietsWeight2016,laviada-molinaEffectsNonnutritiveSweeteners2020,liEffectsNonnutritiveSweeteners2025,ndanukoDietaryPatternsBlood2016,rogersEffectsLowcalorieSweeteners2021,saneeiInfluenceDietaryApproaches2014,shannonMediterraneanDietIncreases2020,cowellEffectsMediterraneanDiet2021,zareEffectSteviaBlood2024,chiavaroliDietaryFiberEffects2015,guptaEffectSolubleFiber2025,jovanovskiShouldViscousFiber2019,khanEffectViscousSoluble2018,postDietaryFiberTreatment2012,silvaFiberIntakeGlycemic2013,soDietaryFiberIntervention2018,streppelDietaryFiberBlood2005,thompsonEffectsIsolatedSoluble2017,wheltonEffectDietaryFiber2005,blochOmega3FattyAcids2012,brownOmega3Omega6Total2019,goldbergMetaanalysisAnalgesicEffects2007,grossoRoleOmega3Fatty2014,huMarineOmega3Supplementation2019,khorshidiEffectOmega3Supplementation2023,kongExplorationOptimizedPortrait2025a,kotwalOmega3Fatty2012,liaoEfficacyOmega3PUFAs2019,rizosAssociationOmega3Fatty2012,serraSupplementationOmega32021,suAssociationUseOmega32018,wangEffectOmega3Fatty2012,cermakProteinSupplementationAugments2012,fingerEffectsProteinSupplementation2015,liaoEffectsProteinSupplementation2017,mortonSystematicReviewMetaanalysis2018,nunesSystematicReviewMetaanalysis2022,orssoEffectsHighproteinSupplementation2024,schoenfeldEffectProteinTiming2013,wycherleyEffectsEnergyrestrictedHighprotein2012,autierVitaminSupplementationTotal2007,bergmanVitaminRespiratoryTract2013,beveridgeEffectVitaminSupplementation2015,bischoff-ferrariFallPreventionSupplemental2009,bischoff-ferrariFracturePreventionVitamin2005,bjelakovicVitaminSupplementationPrevention2014,chenEfficacyVitaminSupplementation2024,jolliffeVitaminSupplementationPrevent2021,juraschekEffectsVitaminSupplementation2012,macphersonMultivitaminmultimineralSupplementationMortality2013,martineauVitaminSupplementationPrevent2017,mayo-wilsonVitaminSupplementsPreventing2011,rakshasbhuvankarVitaminSupplementationVerypreterm2021,reidEffectsVitaminSupplements2014,stocktonEffectVitaminSupplementation2010,yaoVitaminCalciumPrevention2019,zhangAssociationVitaminSupplementation2019,grgicEffectsCaffeineIngestion2019,irwinEffectsAcuteCaffeine2020,politoAcuteEffectCaffeine2016,rayagonzalezAcuteEffectsCaffeine2020,shenEstablishingRelationshipEffect2019,southwardEffectAcuteCaffeine2018,wangEffectsCaffeineIntake2022,caiEffectCoffeeConsumption2012,chenEffectCaffeineIngestion2024,grgicEffectsCaffeineIntake2018,daiAdditionalEffectExercise2025,fernandoEffectRamadanFasting2019,guEffectsIntermittentFasting2022,jahramiSystematicReviewMetaanalysis2020,khalafiEfficacyIntermittentFasting2025,kord-varkanehInfluenceFastingEnergy2020,mengEffectsIntermittentFasting2020,vieiraEffectsAerobicExercise2016,wangEffectsIntermittentFasting2020,xieEffectsTimeRestrictedEating2024,yangEffectEpidemicIntermittent2021,riosleyvrazHealthEffectsUse2022,abdelhamidOmega3FattyAcids2020,milneMetaanalysisProteinEnergy2006}

	\bibliographystyle{plainnat}
	\bibliography{main}
	
\end{document}